%% file: 00_Main.tex
\documentclass[review]{elsarticle}

\usepackage{textgreek}
\usepackage{amsmath}
\usepackage{gensymb}
\usepackage{siunitx}
\usepackage{longtable}
\usepackage{caption}
\usepackage{subcaption}
\usepackage{mathrsfs}
\usepackage{amssymb}
\usepackage{stackengine}
\usepackage{siunitx}

\usepackage{framed} 
\usepackage{multicol} 
\usepackage{nomencl} 

\usepackage{etoolbox}
\renewcommand\nomgroup[1]{%
  \item[\bfseries
  \ifstrequal{#1}{A}{Acronyms}{%
  \ifstrequal{#1}{P}{Superscripts}{
  \ifstrequal{#1}{S}{Subscripts}{
  \ifstrequal{#1}{V}{Variables}}}}%
]}

\newcommand{\nomunit}[1]{%
\renewcommand{\nomentryend}{\hspace*{\fill}\si{#1}}}

\journal{Applied Energy}

\begin{document}

\begin{frontmatter}

\title{Dynamic Modeling and Control of a Two-Reactor Metal Hydride Energy Storage System}

\author[mymainaddress]{Patrick Krane}
\author[mymainaddress]{Austin L. Nash}
\author[mymainaddress]{Davide Ziviani}
\author[mymainaddress]{James E. Braun}
\author[mymainaddress]{Amy M. Marconnet}
\author[mymainaddress]{Neera Jain}

\address[mymainaddress]{School of Mechanical Engineering, Purdue University, West Lafayette, IN}

\begin{abstract}
Metal hydrides have been studied for use in energy storage, hydrogen storage, and air-conditioning (A/C) systems. A common architecture for A/C and energy storage systems is two metal hydride reactors connected to each other so that hydrogen can flow between them, allowing for cyclic use of the hydrogen. This paper presents a nonlinear dynamic model and multivariate control strategy of such a system. Each reactor is modelled as a shell-and-tube heat exchanger connected to a circulating fluid, and a compressor drives hydrogen flow between the reactors. We further develop a linear state-space version of this model integrated with a model predictive controller to determine the fluid mass flow rates and compressor pressure difference required to achieve desired heat transfer rates between the metal hydride and the fluid. A series of case studies demonstrates that this controller can track desired heat transfer rates in each reactor, even in the presence of time-varying  circulating fluid inlet temperatures, thereby enabling the use of a two-reactor system for energy storage or integration with a heat pump.

\end{abstract}


\end{frontmatter}

\makenomenclature
\setlength{\nomitemsep}{-\parskip} 
\begin{multicols}{2}

\nomenclature[Vm]{$m$}{mass \nomunit{\kilogram}}
\nomenclature[Vc]{$c$}{specific heat \nomunit{\joule/\kilogram \cdot \kelvin}}
\nomenclature[Vcp]{$c_{p}$}{constant-pressure specific heat \nomunit{\joule/\kilogram \cdot\kelvin}}
\nomenclature[VT]{$T$}{temperature \nomunit{\kelvin}}
\nomenclature[Vt]{$t$}{time \nomunit{\second}}
\nomenclature[Vr]{$r$}{reaction rate \nomunit{\kilogram \ H / \kilogram \ M \cdot \second}}
\nomenclature[VzD]{$\Delta H$}{enthalpy of reaction \nomunit{\joule/\kilogram \ H}}
\nomenclature[Vze]{$\epsilon$}{effectiveness \nomunit{-}}
\nomenclature[Vm]{$\dot{m}$}{mass flow rate \nomunit{\kilogram / \second}}
\nomenclature[VQ]{$\dot{Q}$}{heat transfer rate \nomunit{\watt}}
\nomenclature[VP]{$P$}{pressure \nomunit{\kilo\pascal}}
\nomenclature[VR{$R$}{gas constant \nomunit{\joule/\kilogram \cdot\kelvin}}
\nomenclature[VV]{$V$}{volume \nomunit{\meter\cubed}}
\nomenclature[Vzpz]{$\phi$}{porosity \nomunit{-}}
\nomenclature[Vw]{$w$}{weight fraction \nomunit{\kilogram \ H / \kilogram \ M}}
\nomenclature[VCA]{$C_{A}$}{Arrhenius pre-exponential factor \nomunit{1 / \second}}
\nomenclature[VEA]{$E_{A}$}{activation energy \nomunit{\joule / \mole \ H}}
\nomenclature[Vzm]{$\mu$}{chemical potential \nomunit{\joule / \mole \ H}}
\nomenclature[VA]{$A$}{phase interaction energy \nomunit{\joule / \mole \ H}}
\nomenclature[VzD]{$\Delta S \degree$}{entropy of reaction \nomunit{\joule / \mole \ H \cdot \kelvin}}
\nomenclature[VzD]{$\Delta H \degree$}{enthalpy of reaction \nomunit{\joule / \mole \ H}}
\nomenclature[VRe]{$Re_{D}$}{Reynolds number \nomunit{-}}
\nomenclature[VPr]{$Pr$}{Prandtl number \nomunit{-}}
\nomenclature[VAS]{$A_{S}$}{surface area \nomunit{\meter \square}}
\nomenclature[VAC]{$A_{c}$}{cross-sectional area \nomunit{\meter \square}}
\nomenclature[Vh]{$h$}{heat transfer coefficient \nomunit{\watt/\meter\square \ \cdot \kelvin}}
\nomenclature[Vzr]{$\rho$}{density \nomunit{\kilogram / \meter \cubed}}
\nomenclature[VKl]{$K_{loss}$}{loss coefficient \nomunit{-}}
\nomenclature[VA]{$\mathbf{A}$}{matrix in state-space eqn.}
\nomenclature[VB]{$\mathbf{B}$}{matrix in state-space eqn.}
\nomenclature[VC]{$\mathbf{C}$}{matrix in state-space eqn.}
\nomenclature[VD]{$\mathbf{D}$}{matrix in state-space eqn.}
\nomenclature[Vx]{$x$}{state variables}
\nomenclature[Vu]{$u$}{input variable}
\nomenclature[VzD]{$\Delta P$}{change in pressure \nomunit{\kilo \pascal}}
\nomenclature[Vf]{$f$}{state variable rate of change}
\nomenclature[Vg]{$g$}{output variable rate of change}
\nomenclature[VD]{$D$}{diameter \nomunit{\meter}}
\nomenclature[Vk]{$k$}{thermal conductivity \nomunit{\watt / \meter \cdot \kelvin}}
\nomenclature[Shy]{$hyd$}{hydride}
\nomenclature[SH]{$H$}{hydrogen}
\nomenclature[Sr]{$r$}{reactor}
\nomenclature[Swg]{$wg$}{water glycol}
\nomenclature[Sin]{$in$}{inlet}
\nomenclature[Ssh]{$shell$}{shell}
\nomenclature[Seq]{$eq$}{equilibrium}
\nomenclature[Smax]{$max$}{maximum}
\nomenclature[Sza]{$\alpha,0$}{$\alpha$ phase boundary}
\nomenclature[Szb]{$\beta,0$}{$\beta$ phase boundary}
\nomenclature[Sc]{$c$}{critical}
\nomenclature[SA]{$A$}{MmNi\textsubscript{4.5}Cr\textsubscript{0.5} reactor}
\nomenclature[SB]{$B$}{LaNi\textsubscript{5} reactor}
\nomenclature[Sout]{$out$}{outlet}
\nomenclature[S ]{$\rightarrow$}{to}
\nomenclature[SHl]{$H\ line$}{hydrogen line}
\nomenclature[S0]{$0$}{linearization point}
\nomenclature[Satm]{$atm$}{atmosphere}
\nomenclature[Sm]{$m$}{absorbing reactor}
\nomenclature[Sn]{$n$}{desorbing reactor}
\nomenclature[Sabs]{$abs$}{absorption}
\nomenclature[Sdes]{$des$}{desorption}
\nomenclature[SM]{$M$}{metal}
\nomenclature[Si]{$i$}{initial}
\nomenclature[Stube]{$tube$}{tube}
\printnomenclature
\end{multicols}


\input{01_Introduction}

\input{02a_Modeling}

\input{02b_Model_Validation}

\input{03a_Control_Design}

\input{03b_Controller_Validation}

\input{04_Conclusion}

\input{05_Acknowledgements}

\bigskip\bigskip
 \bibliographystyle{model1-num-names}
 \bibliography{references}
 
\input{06_Appendix}
 
\end{document}

%% file: 01_Introduction.tex
\section{Introduction}


\par Metal hydrides are compounds that form when certain metals react with hydrogen. The hydriding reaction is reversible: the metal absorbs hydrogen exothermically at certain temperatures and pressures and desorbs hydrogen endothermically under other conditions. The search for new energy storage technologies has led to investigation of metal hydrides for this purpose, including using hydrides as a compact form of hydrogen storage for fuel cells \cite{Tange2011ExperimentalSystem,Khayrullina2018AirApplications,Aruna2020ModelingBed,Cho2013DynamicTank,Panos2010DynamicTank,Georgiadis2009DesignStorage} and for energy storage integrated with concentrated solar power generation \cite{Corgnale2016MetalPlants,Feng2019OptimumStorage,Sheppard2016MetalStorage,SatyaSekhar2012TestsSystem,Nyamsi2018SelectionTargets}. Systems with multiple metal hydride reactors have also been evaluated for heating and cooling applications, where the hydride reactors are used in place of a heat pump \cite{Kang1995ThermalConditioning,Muthukumar2010MetalReview,Nagel1984OperatingAir,Magnetto2006ASystem,Satheesh2010PerformancePump,Satheesh2010SimulationPump,Nie2011MetalDriving}. A common design for metal hydride systems in energy storage, heating, and cooling applications is to have a pair of metal hydride reactors connected so that hydrogen can flow between them. In other words, as hydrogen is desorbed in one reactor, it flows to the other reactor, which absorbs it. 


\par When used for energy storage, the two-reactor system does not run continuously but instead absorbs heat from an external source during one half-cycle and releases it during the other half-cycle \cite{Corgnale2016MetalPlants,Sheppard2016MetalStorage,Nyamsi2018SelectionTargets,Manickam2019FutureHydrides,Ward2016TechnicalSystems}. In this application, only one reactor is directly used for energy storage, while the other reactor is used to store the hydrogen released by that reactor \cite{Corgnale2016MetalPlants}. In general, two hydride beds are not required for energy storage systems, since hydrogen released by the metal hydride can be compressed, stored, and released from a pressure vessel \cite{Feng2019OptimumStorage,Sheppard2016MetalStorage,SatyaSekhar2012TestsSystem}. However, a two-reactor design is commonly studied since it allows for more compact hydrogen storage without the need to compress hydrogen to high pressures. The two-reactor metal hydride system is also used in single-stage single-effect air conditioning systems, where the hydride reactions are controlled by changing the heat sources and sinks connected to the reactors \cite{Muthukumar2010MetalReview}. The pair of hydride reactors undergo a two-step cycle. In the first (charging) step, the hydride in one reactor is heated by a high-temperature heat source and desorbs hydrogen that flows to the other reactor. Heat is released as the hydrogen is absorbed in the second reactor, and in turn, the heat is transferred to a heat sink at an intermediate temperature. In the second (discharging) step, the first reactor is connected to the intermediate-temperature heat sink while the second reactor is connected to the environment to which the cooling load needs to be delivered. It absorbs heat from this location, providing the cooling load, as it desorbs hydrogen that then flows to the other reactor \cite{Muthukumar2010MetalReview}. In order for this system to operate as intended, the metal hydrides have to be selected so that in each step, changing the heat source or sink connected to the reactor results in a temperature shift large enough to drive the desorption or absorption reaction \cite{Satheesh2010PerformancePump} respectively.

\par If this system design is used for continuous cooling instead of energy storage, a four-reactor system is necessary, with one reactor pair in each step of the cycle at any given time \cite{Kang1995ThermalConditioning,Muthukumar2010MetalReview,Nagel1984OperatingAir}. An alternative design, which uses a \emph{single pair} of connected reactors, provides a continuous cooling load by using a compressor to move hydrogen between the reactors, thus driving the reaction in each reactor by changing the pressure, instead of the temperature
\cite{Magnetto2006ASystem}. This system design requires additional power for operating the compressor, but allows for more flexibility in choosing the specific hydrides used in the reactors.  However, to realize the benefits of this system, specifically to guarantee specific heat transfer rates between the hydride reactors and a circulating fluid, a real-time embedded control strategy for the system is needed.

\par Numerous models of the heat and mass transfer in metal hydride reactors have been developed. The majority of these focus on a single metal hydride reactor, including 1-D models \cite{Choi1990HeatApplications,Abdin2018One-dimensionalMatlabSimulink} and 2-D models \cite{Nakagawa2000NumericalBed,Jemni1999ExperimentalReactor,Brown2008AccurateTank}. More computationally-intensive 3-D models of a hydride reactor, first developed by \citet{Aldas2002ABed}, have been used to analyze the performance benefits of complicated system geometries, such as the addition of fins \cite{Askri2009HeatMmNi4.6Fe0.4} or a tapered bed structure \cite{Feng2019OptimumStorage} to improve heat transfer. With respect to two-reactor systems, modeling has been more limited;~\citet{Kang1995ThermalConditioning} used a 1-D model to examine a metal hydride air conditioning system and \citet{Nyamsi2018SelectionTargets} used a 3-D model to examine a pair of hydride reactors used for energy storage. Moreover, neither of these models considered a pair of hydride reactors with an  auxiliary hydrogen compressor.

Similarly, control strategies for metal hydride reactors have focused on single-reactor systems used for hydrogen storage; relevant literature is summarized in Table~\ref{tab:lit_rev}. Since these papers are focused on a single-reactor system for hydrogen storage, the output variable of the controllers is either the hydrogen flow rate or the temperature or pressure of the reactor, with the latter generally being used because of its influence on the reaction rate in the reactor. For instance, while
\citet{Georgiadis2009DesignStorage} and \citet{Panos2010DynamicTank} use outlet temperature as the output variable of the controller, they do so in order to achieve the outlet temperature that will result in the optimal hydrogen release rate. Similarly, the input variables considered in all of these papers, except for \citet{Cho2013DynamicTank}, are those used to control the heat source used to drive the desorption in the reactor, whether that be the flow rate \cite{Georgiadis2009DesignStorage,Panos2010DynamicTank} or temperature \cite{Aruna2020ModelingBed} of a heating fluid, or the voltage of a thermoeletric heater \cite{Nuchkrua2013Neuro-fuzzyReactor}. While similar inputs would be considered for a case where the hydride reactor was being used for energy storage or air conditioning, the output variable for such cases would be the heat transfer to or from the reactor, which is not considered by any of these controllers. 

\begin{table}[!htb]
\caption{Summary of the controller types and input and output variables that have previously been considered for single reactor metal hydride systems.} 
\begin{center}
  \resizebox{\hsize}{!}{\begin{tabular}{p{25 mm} p{30 mm} p{25 mm} p{25 mm}}
  \noalign{\vskip -1.5mm}
  \hline
  \noalign{\vskip 1mm}
  \textbf{Author(s)} & \textbf{Controller} & \textbf{Control} \newline \textbf{Variable} & \textbf{Output} \newline \textbf{Variable}  \\
  \noalign{\vskip 1mm}
  \hline
  \noalign{\vskip 1mm}
    \citet{Georgiadis2009DesignStorage} 2009 & Multi-parametric (mp) MPC & Heating Fluid Flow Rate &  Outlet \newline Temperature \\
    \citet{Panos2010DynamicTank} 2010 & mp-MPC & Heating Fluid Flow Rate &  Outlet \newline Temperature \\
    \citet{Cho2013DynamicTank} 2013 & PID & Discharge Valve Opening & Hydrogen Flow Rate \\
    \citet{Nuchkrua2013Neuro-fuzzyReactor} 2013 & Neuro-fuzzy PID & Thermoelectric Voltage & Reactor \newline Temperature \\
    \citet{Aruna2020ModelingBed} 2020  & Fuzzy PID & Heating \newline Temperature & Reactor \newline Pressure \\
  \noalign{\vskip 1mm}
  \hline
  \end{tabular}}
\end{center}
\label{tab:lit_rev}
\end{table}

\par Different models of the hydride system are also used for controller design. For instance, \citet{Georgiadis2009DesignStorage} use a 2-D model to determine the temperature and pressure variations within the reactor, whereas \citet{Cho2013DynamicTank} use a model that neglects the effect of these variations. The linear models used for control synthesis are obtained using model identification techniques and simulation data from the system model \cite{Georgiadis2009DesignStorage,Panos2010DynamicTank,Aruna2020ModelingBed}. \citet{Aruna2020ModelingBed} examine different methods of converting their nonlinear model into a linear model for use in a controller, and find the best results when using a Box-Jenkins model. All of the methods compared are black-box methods \cite{Aruna2020ModelingBed}; that is, they develop the linear model using only input and output data rather than deriving such a model from first principles. This poses a disadvantage from the perspective of control design because it limits the operation of the controller to a small region around the linearization point.

With respect to two-reactor systems, experimental demonstrations \cite{Nagel1984OperatingAir,Magnetto2006ASystem} used logic-based controllers to operate the system, but did not involve the design of real-time controllers that could achieve specific heat transfer rates at specific times, as would be desired for an energy storage system which is charged and discharged over a longer period of time and may not be charged or discharged continuously. These experimental systems have also used either heat transfer from a fluid or a compressor to drive the system operation, but have not examined how to control a system that uses both. In summary, previous work on control strategies for metal hydride reactors has not considered real-time control for a two-reactor system using both temperature and pressure to drive operation, nor control using a linear model derived from the system dynamics rather than from a black-box model.

\par Therefore, in this paper, we present a dynamic model for a metal hydride energy storage system along with a model predictive control strategy for tracking the desired heat transfer rates in each reactor of a two-reactor metal hydride system. Specifically, in Section~\ref{sec:model}, we present the dynamic model of the metal hydride energy storage system including two metal hydride reactors and a compressor to drive hydrogen flow. Then, we derive the linear state-space form of the model using first principles. The linearized model is validated against the nonlinear one, and we highlight how the ability to update the linearized model enables better agreement between the linear and nonlinear models. The resulting model is then used for control synthesis. Because the linearized model is physics-based and fully parameterized, it can be relinearized around any operating condition in real time, enabling it to be used beyond the nominal operating condition. In Section~\ref{sec:control}, we present the model predictive control (MPC) design for determining the mass flow rates of the circulating fluid and the pressure difference between the hydrogen reactors needed to meet the demanded heat transfer rate in each reactor. Finally, in Section~\ref{sec:results}, we demonstrate the performance of the controller in simulation with a series of case studies. The results indicate that the controller can achieve desired heat transfer rates between hydride beds through control of the mass flow rates of the circulating fluids and the compressor pressure. This controller enables the use of a two-reactor system for energy storage and/or integration with a heat pump.   

%% file: 02a_Modeling.tex
\section{Model Description}
\label{sec:model}
\begin{figure}[!htb]
\centering
\begin{subfigure}{0.9\textwidth}
\includegraphics[width=\linewidth]{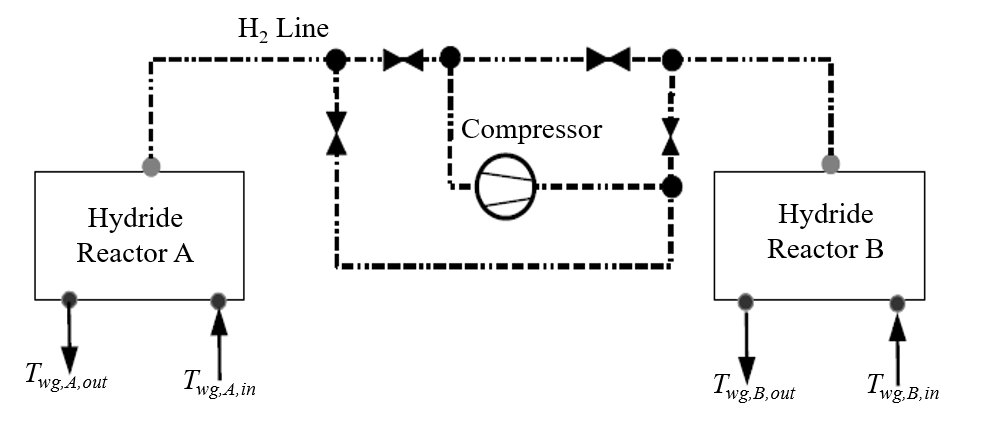}
\caption{Two-reactor metal hydride energy storage system}
\label{fig:Hyd_Schematic_reactors}
\end{subfigure}
\begin{subfigure}{0.9\textwidth}
\includegraphics[width=\linewidth]{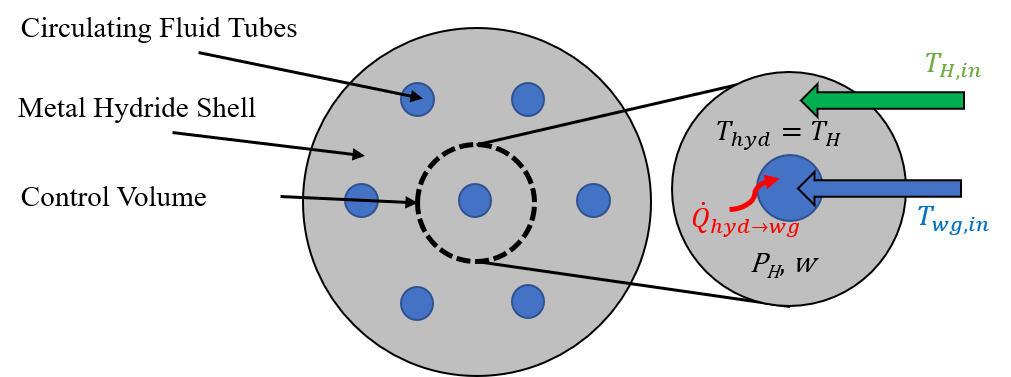}
\caption{Cross-sectional view of a reactor. Seven tubes are shown for clarity, while the proposed system has 400 tubes within the shell.}
\label{fig:Hyd_Schematic_shellandtube}
\end{subfigure}
\caption{Schematic of the metal hydride energy storage system. (a) Two metal hydride reactors are connected so that hydrogen can flow between them. A compressor drives hydrogen flow when the thermally-driven pressure gradient is unfavorable. (b) Each reactor is a shell-and-tube heat exchanger with multiple tubes for the circulating fluid flow and a single shell filled with metal hydride powder. To simplify the heat exchanger model, the analyzed control volume consists of a single tube and the surrounding metal hydride as shown.}
\label{fig:Hyd_Schematic}
\end{figure}

Here we consider a thermal energy storage system consisting of two interconnected metal-hydride reactors with a compressor to drive hydrogen flow between them, as shown in Figure~\ref{fig:Hyd_Schematic_reactors}. Each reactor is modeled as a shell-and-tube heat exchanger (depicted in Figure~\ref{fig:Hyd_Schematic_shellandtube}) absorbing heat from, and releasing heat to, a circulating fluid that interacts with a heat pump. We first derive a dynamic, nonlinear thermodynamic model of the two-reactor system using first principles. We then linearize the model to obtain a form suitable for control synthesis, and validate the linear model against the nonlinear one. 

\subsection{System Model}
\par Each metal hydride reactor is modeled as a shell-and-tube heat exchanger. The shell is filled with a porous bed of metal hydride through which hydrogen flows. Water glycol circulates in the tubes and exchanges heat with the hydride bed. The heat exchanger is modelled using a control volume consisting of one tube and the shell surrounding it, as shown in Figure~\ref{fig:Hyd_Schematic_shellandtube}, with each other tube and surrounding shell assumed to be the same by symmetry. The perimeter of the heat exchanger is assumed to be insulated, so heat losses from the heat exchanger to ambient are neglected.

\par Within the control volume, it is assumed that the hydrogen in the reactors is an ideal gas, that the hydrogen and hydride in the reactor are in thermal equilibrium ($T_{hyd}=T_{H}$), and that all spatial variation in temperature and pressure are negligible. Given these assumptions, the energy balance for a single control volume is given by
\begin{align}
\left (m_{hyd}c_{hyd} + m_{H}c_{p,H}  \right )\frac{dT_{hyd}}{dt} = rm_{hyd}\Delta H + \epsilon \dot{m}_{wg}c_{wg}\left ( T_{wg,in} - T_{hyd} \right) \notag \\
+ \dot{m}_{H,in}c_{p,H}\left ( T_{H,in} - T_{hyd} \right),  \label{eq:ebal}
\end{align} 
where $m_{hyd}$ and $m_H$ are the mass of the hydride and the hydrogen, respectively; $c_{hyd}$, $c_{p,H}$, and $c_{wg}$ are the specific heat of the hydride, hydrogen, and water glycol, respectively; $T_{hyd}$ is the temperature of the metal hydride in the reactor, $r$ is the reaction rate, $\Delta H$ is the heat of reaction, $\epsilon$ is the heat exchanger effectiveness, $\dot{m}_{wg}$ and $\dot{m}_{H,in}$ are the mass flow rates of water glycol and hydrogen into the reactor, and $T_{wg,in}$ and $T_{H,in}$ are the inlet temperatures of water glycol and hydrogen, respectively. The mass balance for the hydrogen gas is given by
\begin{equation}\label{eq:mbal}
\frac{dm_{H}}{dt} =\ rm_{hyd} + \dot{m}_{H,in}.
\end{equation}
Using the ideal gas law, the pressure of the hydrogen gas, $P_{H}$, can be obtained from the mass of hydrogen as
\begin{equation}\label{eq:igas}
P_{H} =\ \frac{m_{H}R_{H}T_{H}}{V_{H}}.
\end{equation}
In this equation, $R_{H}$ is the gas constant for hydrogen. The volume of hydrogen ($V_{H}$) can be found using the porosity of the metal hydride bed ($\phi$) and the total volume of the hydride section of the heat exchanger ($V_{shell}$) according to
\begin{equation}\label{eq:porosity}
V_{H}=\ \phi V_{shell}.
\end{equation}
Using Equations~\ref{eq:igas} and~\ref{eq:porosity} and the thermal equilibrium assumption ($T_{hyd}$ =$T_{H}$), the mass balance can be re-written in terms of pressure as
\begin{equation}\label{eq:pbal}
\frac{dP_{H}}{dt} =\ \frac{RT_{hyd}}{\phi V_{shell}}\left (rm_{hyd} + \dot{m}_{H,in}  \right ).
\end{equation}

The reaction rate, $r$, is the rate of change of the weight fraction of hydrogen stored in the metal hydride, $w$. The equations for the reaction rate and equilibrium pressure, $P_{eq}$, are taken from Voskuilen
\cite{Voskuilen2014ASystems}.
The reaction rate is given by
\begin{equation}\label{eq:rrate}
r =\frac{dw}{dt} = \ \left\{\begin{matrix} C_{A}e^{-\frac{E_{A}}{RT_{hyd}}}ln\left ( \frac{P}{P_{eq}} \right )\left ( w_{max} - w \right ), & P > P_{eq,abs} \\ 0, & P_{eq,des} < P < P_{eq,abs}
\\ C_{A}e^{-\frac{E_{A}}{RT_{hyd}}}ln\left ( \frac{P}{P_{eq}} \right )w, & P < P_{eq,des} .
\end{matrix}\right.
\end{equation}
In these equations, $C_{A}$ and $E_{A}$ are constant material properties of the metal hydride, $w_{max}$ is the maximum weight fraction of the metal hydride, and the equilibrium pressures are given by
\begin{equation}\label{eq:Peq}
P_{eq}=\ P_{atm} e ^{\frac{\mu\left ( w,T \right )}{RT}},
\end{equation}
where the chemical potential, $\mu(w,T)$, is defined as
\begin{equation}\label{eq:mu}
\mu =\ \left\{\begin{matrix}\mu_{\alpha,0} + 2RT_{c}\left (1-2\frac{w}{w_{max}}  \right ) + RT ln \left (  \frac{w}{w_{max}-w}\right ), & \ w < w_{\alpha,0}\\ 
\Delta H\degree - T \Delta S\degree + A \left ( \frac{w}{w_{max}} - 0.5 \right ), &  w_{\alpha,0} < w < w_{\beta,0}\\
\mu_{\beta,0} + 2RT_{c}\left (1-2\frac{w}{w_{max}}  \right )+ RT ln \left (  \frac{w}{w_{max}-w}\right ), &  w > w_{\beta,0}
\end{matrix}\right.
\end{equation}
\noindent In this equation, $\alpha$ and $\beta$ are different phases of the metal hydride. The material properties $\Delta H\degree$ and $\Delta S\degree$ have different values for absorption and desorption, resulting in different equilibrium pressures for absorption and desorption.

\par In Equations~\ref{eq:Peq} and~\ref{eq:mu}, $P_{atm}$ is atmospheric pressure, $w_{\alpha,0}$ and $w_{\beta,0}$ are the weight fractions at which the hydride phase changes, $\mu_{\alpha,0}$ and $\mu_{\beta,0}$ are the chemical potential at $w_{\alpha,0}$ and $w_{\beta,0}$, respectively, and $T_{c}$ and $A$ are the critical temperature and phase interaction energy of the metal hydride, respectively. In the energy balance, the heat exchanger effectiveness is calculated using the expression
\begin{equation}\label{eq:hxeff}
\epsilon =\ 1 - e^{-\frac{hA_{S}}{\dot{m}_{wg}c_{wg}}},
\end{equation}
where $h$ is the heat transfer coefficient and $A_s$ is the surface area of the tube. The heat transfer coefficient is calculated from the Reynolds number, $Re_{D}$, Prandtl number, $Pr$, and thermal conductivity, $k$, of the fluid using the Dittus-Boelter correlation for fully-developed turbulent flow in a circular tube
\cite{Incropera2011FundamentalsEdition}:
\begin{equation}\label{eq:hconv}
Nu =\ \frac{hD_{tube}}{k_{wg}} =\ 0.023 Re_{D}^{4/5} Pr^{n}.
\end{equation}

\noindent Then, the outlet temperature of the circulating fluid, $T_{wg,out}$ can be calculated as
\begin{equation}\label{eq:Tout}
    T_{wg,out}=\ T_{wg,in} + \epsilon \left ( T_{hyd} - T_{wg,in} \right ).
\end{equation}
Thus, the heat transfer between the hydride and the water glycol, $\dot{Q}_{hyd \rightarrow wg}$, is given by
\begin{equation}\label{eq:Qdot}
    \dot{Q}_{hyd \rightarrow wg}=\ \dot{m}c_{wg}\left ( T_{wg,out}-T_{wg,in} \right ) = \epsilon \dot{m}c_{wg}\left ( T_{hyd}-T_{wg,in} \right ).
\end{equation}

\noindent Together, Equations \ref{eq:ebal} through \ref{eq:Qdot} are used to solve for the state of each reactor. The reactors are coupled through the mass flow rate of hydrogen into the reactor, which is calculated for one reactor as
\begin{equation}\label{eq:mdotH_BtoA}
\dot{m}_{H,A}= \left\{\begin{matrix} \ A_{c, H line}\sqrt{\frac{2 \rho_{H} \left ( P_{B} + \Delta P_{comp} - P_{A} \right )}{K_{loss}}}, & \text{flow from B to A}\\
\ -A_{c, H line}\sqrt{\frac{2 \rho_{H} \left ( P_{A} + \Delta P_{comp} - P_{B} \right )}{K_{loss}}}, & \text{flow from A to B}
\end{matrix}\right.
\end{equation}
and for the other reactor as
\begin{equation}\label{eq:mdotH_B}
\dot{m}_{H,B}=\ - \dot{m}_{H,A}.
\end{equation}
In these equations, $A_{c,H line}$ is the cross-sectional area of the hydrogen line connecting the reactors, $\rho_{H}$ is the density of hydrogen, $\Delta P_{comp}$ is the pressure difference created by the compressor, and $K_{loss}$ is the loss coefficient for flow from one reactor to the other. As shown in Figure~\ref{fig:Hyd_Schematic}, the compressor can drive flow in either direction, depending on which valves are open. 

\subsection{Linear State-Space Model}
\par While the governing dynamics of the two-reactor hydride system are nonlinear, the derived model is not well suited for control algorithm synthesis. An alternative approach is to linearize the dynamics about a nominal operating condition and synthesize a controller based on this linearized model.  To that end, we linearize the first principles model derived in the previous subsection using a standard state-space representation as given by Equations \eqref{eq:stspc_x} and \eqref{eq:stspc_y} \cite{Ogata2010ModernEngineering}.
\begin{equation} \label{eq:stspc_x}
f(x,u)=\ \frac{dx}{dt}=\ \mathbf{A}\left ( x - x_{0} \right ) + \mathbf{B}\left ( u - u_{0} \right ) + f(x_{0},u_{0})
\end{equation}

\begin{equation} \label{eq:stspc_y}
g(x,u)=\ y =\ \mathbf{C}\left ( x - x_{0} \right ) + \mathbf{D}\left ( u - u_{0} \right ) + g(x_{0},u_{0})
\end{equation}

\par Here, $x$ is the standard state-space notation for the dynamic state vector of the system, $u$ is the control input vector and $y$ is the output vector. For the case in which one reactor, \textit{m}, is absorbing hydrogen while the other, \textit{n}, is desorbing, the state, output, and input variables are defined as follows:
\begin{equation}\label{eq:def_x}
x=\ \begin{bmatrix}T_{hyd,m}
\ P_{H,m}
\ w_{m}
\ T_{n}
\ P_{H,n}
\ w_{hyd,n}
\end{bmatrix}^{T} \enspace ,
\end{equation}

\begin{equation}\label{eq:def_y}
y=\ \begin{bmatrix}\dot{Q}_{hyd \rightarrow wg,m}
\ \dot{Q}_{hyd \rightarrow wg,n}
\end{bmatrix}^{T}\enspace ,
\end{equation}

\begin{equation}\label{eq:def_u}
u=\ \begin{bmatrix}\dot{m}_{wg,m}
\ \dot{m}_{wg,n}
\ \Delta P_{comp}
\ T_{wg,in,m}
\ T_{wg,in,n}
\end{bmatrix}^{T} \enspace .
\end{equation}

Note that in these equations, the reactor where \textit{absorption} occurs is always referred to as reactor $m$, and the reactor where \textit{desorption} occurs as reactor $n$, even though these are different reactors depending on the operating mode of the system. For brevity, the complete linearized model equations are not presented here.  However, the authors have made the linearized model code available to the public via Github \cite{Krane2021HydrideModel}.


%% file: 02b_Model_Validation.tex
\section{Linearized Model Validation}
\label{sec:modelval}

In this section, we validate the linearized model against the nonlinear one.  We first describe the specific parameterization of the model based on chosen material properties and system dimensions, followed by a comparison of the linearized and nonlinear models through numerical simulations.

\subsection{Model Parameterization}
Here we consider MmNi\textsubscript{4.5}Cr\textsubscript{0.5} as the material in reactor A and  LaNi\textsubscript{5} as the material in reactor B. These materials were selected based on the operating temperatures of an air conditioning system with which this storage system could ultimately be integrated. The properties of these materials are taken from the toolbox developed by \citet{Voskuilen2014ASystems} and are summarized in Table \ref{tab:hyd_props}. By deriving the nonlinear model, and its linearization, from first principles, the model can easily be parameterized for any choice of metal hydrides.

\begin{table}[!htb]
\caption{Material properties of  MmNi\textsubscript{4.5}Cr\textsubscript{0.5} (Reactor A) and LaNi\textsubscript{5} (Reactor B). Since the density and specific heat of MmNi\textsubscript{4.5}Cr\textsubscript{0.5} are not reported in literature, they are estimated from values for this class of alloys  \cite{Voskuilen2014ASystems}.} 

\begin{center}
  \begin{tabular}{l c c c c}
  \noalign{\vskip -1.5mm}
  \hline
  \noalign{\vskip 1mm}
  \small{$\textbf{\text{Property}}$} & \small{$\textbf{\text{Symbol}}$} & \small{$\textbf{\text{Units}}$} & \small{$\textbf{\text{LaNi\textsubscript{5}}}$} & \small{$\textbf{\text{MmNi\textsubscript{4.5}Cr\textsubscript{0.5}}}$}\\
  \noalign{\vskip 1mm}
  \hline
  \noalign{\vskip 1mm}
    \small{Density} & \small{$\rho$} & \small{kg/$\text{m}^3$}  & \small{8300} & \small{8200}\\
    \small{Specific Heat} & \small{$c$} & \small{J/kg$\cdot$K}  & \small{355} & \small{419}\\
    \small{Enthalpy of Reaction (abs.)} & \small{$\Delta H_{a}$} & \small{MJ/kg M} & \small{15.46} & \small{11.67}\\ 
    \small{Enthalpy of Reaction (des.)} & \small{$\Delta H_{d}$} & \small{MJ/kg M} & \small{15.95} & \small{12.65}\\
    \small{Maximum Weight Fraction} & \small{$w_{max}$} & \small{kg H/kg M} & \small{0.0151} & \small{0.0121} \\
  \noalign{\vskip 1mm}
  \hline
  \end{tabular}
\end{center}
\label{tab:hyd_props}
\end{table}

\par Each hydride reactor, modeled as a shell-and-tube heat exchanger, has the dimensions described in Table \ref{tab:hx_dim}. The different lengths of each hydride reactor are due to their different storage capacities and thus different volumes of hydride required. The pipe for transferring hydrogen between the reactors has a diameter of 1 cm and an assumed overall loss coefficient ($K_{loss}$) of 40.

\begin{table}[!htb]
\caption{Dimensions of the shell-and-tube heat exchangers for the hydride reactors. Only the length varies between the two reactors. Each tube and its surrounding shell represent 1 control volume within the system shown in Figure~\ref{fig:Hyd_Schematic}.} 
\begin{center}
  \begin{tabular}{l c c}
  \noalign{\vskip -1.5mm}
  \hline
  \noalign{\vskip 1mm}
  $\textbf{\text{Property}}$ & $\textbf{\text{Value}}$ & $\textbf{\text{Units}}$ \\
  \noalign{\vskip 1mm}
  \hline
  \noalign{\vskip 1mm}
    Tube Diameter & 4 & mm \\
    Shell Diameter & 7 & mm \\
    Number of Tubes & 400 & - \\ 
    Length of Reactor A & 1.77 & m \\
    Length of Reactor B & 1.54 & m \\
  \noalign{\vskip 1mm}
  \hline
  \end{tabular}
\end{center}
\label{tab:hx_dim}
\end{table}

\subsection{Simulation Results}

\par
\begin{figure}[!htb]
     \centering
     \begin{subfigure}{0.49\textwidth}
         \centering
         \includegraphics[width=\textwidth]{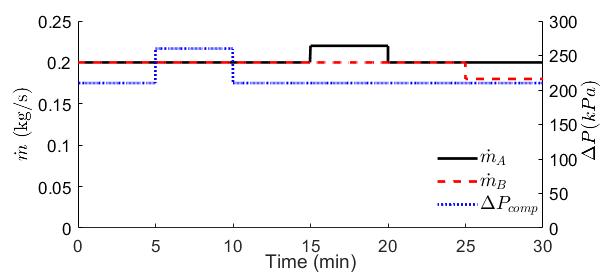}
         \caption{Case 1}
         \label{fig:dP_case1}
     \end{subfigure}
     \begin{subfigure}{0.49\textwidth}
         \centering
         \includegraphics[width=\textwidth]{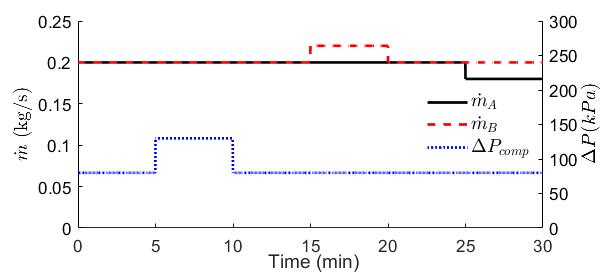}
         \caption{Case 2}
         \label{fig:mdot_case1}
     \end{subfigure}
     \caption{Input circulating fluid mass flow rate and compressor pressure difference for (a) Case 1  and (b) Case 2.}
        \label{fig:Case1&2_inputs}
\end{figure}

\begin{figure}[!htb]
     \centering
     \begin{subfigure}{0.49\textwidth}
         \centering
         \includegraphics[width=\textwidth]{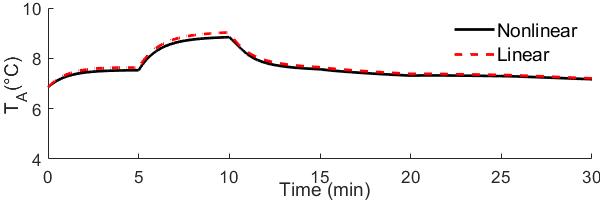}
         \caption{Hydride Temperature in Reactor A}
         \label{fig:tempA_case1}
         \vspace*{2.5mm}
     \end{subfigure}
     \begin{subfigure}{0.49\textwidth}
         \centering
         \includegraphics[width=\textwidth]{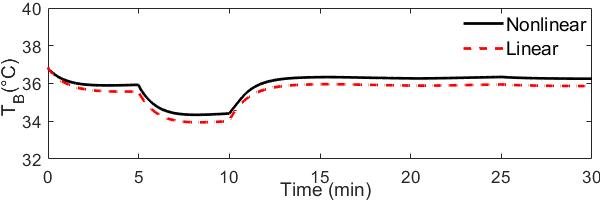}
         \caption{Hydride Temperature in Reactor B}
         \label{fig:tempB_case1}
         \vspace*{2.5mm}
     \end{subfigure}
     \begin{subfigure}{0.49\textwidth}
         \centering
         \includegraphics[width=\textwidth]{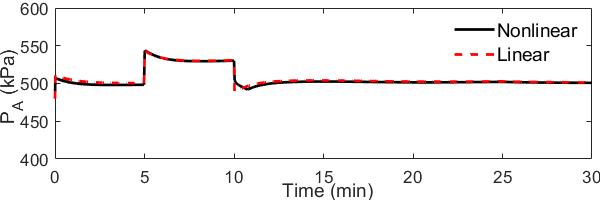}
         \caption{Hydrogen Pressure in Reactor A}
         \label{fig:pressA_case1}
         \vspace*{2.5mm}
     \end{subfigure}
     \begin{subfigure}{0.49\textwidth}
         \centering
         \includegraphics[width=\textwidth]{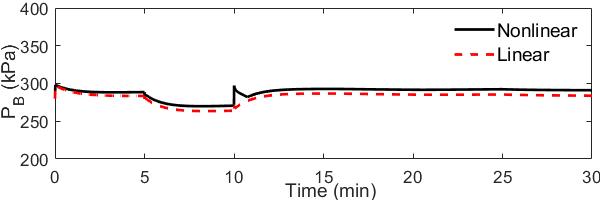}
         \caption{Hydrogen Pressure in Reactor B}
         \label{fig:pressB_case1}
         \vspace*{2.5mm}
     \end{subfigure}
     \begin{subfigure}{0.49\textwidth}
         \centering
         \includegraphics[width=\textwidth]{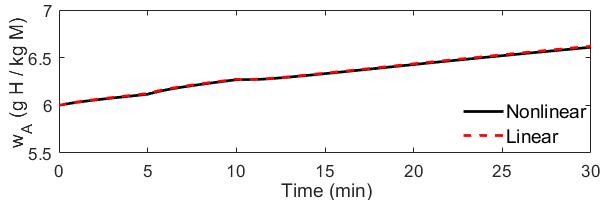}
         \caption{Hydrogen Weight Fraction in Reactor A}
         \label{fig:weightA_case1}
     \end{subfigure}
     \begin{subfigure}{0.49\textwidth}
         \centering
         \includegraphics[width=\textwidth]{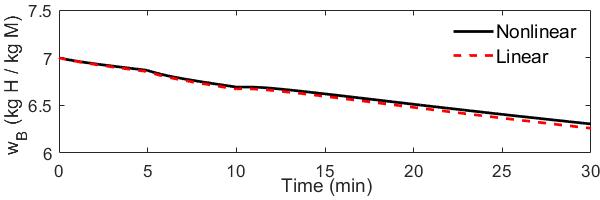}
         \caption{Hydrogen Weight Fraction in Reactor B}
         \label{fig:weightB_case1}
     \end{subfigure}
        \caption{Comparison of the dynamic state variables for Reactors A and B between the linear model (dashed lines) and the nonlinear model (solid lines) for Case 1 (hydrogen flowing from reactor A to reactor B).}
        \label{fig:Case1_lin_perf}
\end{figure}

\begin{figure}[!htb]
     \centering
     \begin{subfigure}{0.49\textwidth}
         \centering
         \includegraphics[width=\textwidth]{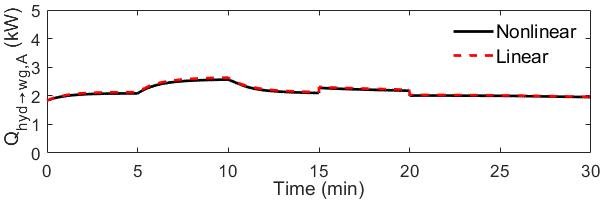}
         \caption{Reactor A}
         \label{fig:heatA_case1}
     \end{subfigure}
     \begin{subfigure}{0.49\textwidth}
         \centering
         \includegraphics[width=\textwidth]{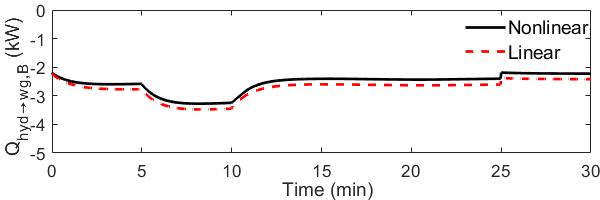}
         \caption{Reactor B}
         \label{fig:heatB_case1}
     \end{subfigure}
     \caption{Comparison of the heat transfer rates associated with Reactors A and B between the linear model (dashed lines) and the nonlinear model (solid lines) for Case 1 (hydrogen flowing from reactor A to reactor B).}
        \label{fig:Case1_lin_perf_heat}
\end{figure}

\begin{figure}[!htb]
     \centering
     \begin{subfigure}{0.49\textwidth}
         \centering
         \includegraphics[width=\textwidth]{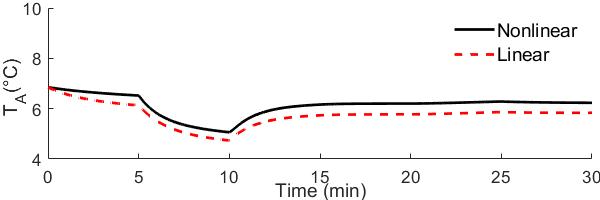}
         \caption{Hydride Temperature in Reactor A}
         \label{fig:tempA_case2}
         \vspace*{2.5mm}
     \end{subfigure}
     \begin{subfigure}{0.49\textwidth}
         \centering
         \includegraphics[width=\textwidth]{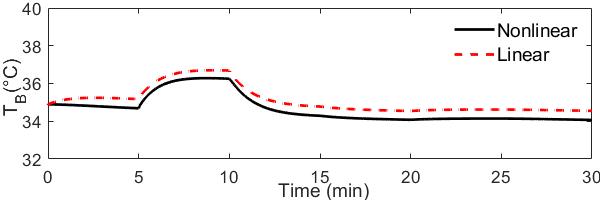}
         \caption{Hydride Temperature in Reactor B}
         \label{fig:tempB_case2}
         \vspace*{2.5mm}
     \end{subfigure}
     \begin{subfigure}{0.49\textwidth}
         \centering
         \includegraphics[width=\textwidth]{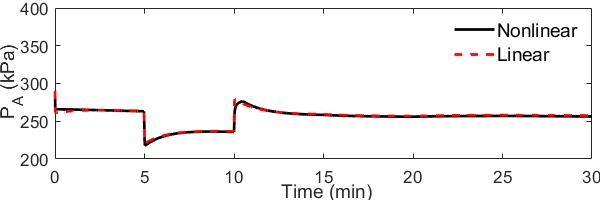}
         \caption{Hydrogen Pressure in Reactor A}
         \label{fig:pressA_case2}
         \vspace*{2.5mm}
     \end{subfigure}
     \begin{subfigure}{0.49\textwidth}
         \centering
         \includegraphics[width=\textwidth]{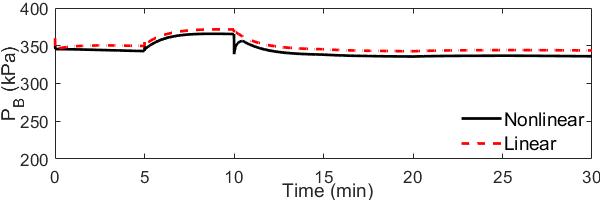}
         \caption{Hydrogen Pressure in Reactor B}
         \label{fig:pressB_case2}
         \vspace*{2.5mm}
     \end{subfigure}
     \begin{subfigure}{0.49\textwidth}
         \centering
         \includegraphics[width=\textwidth]{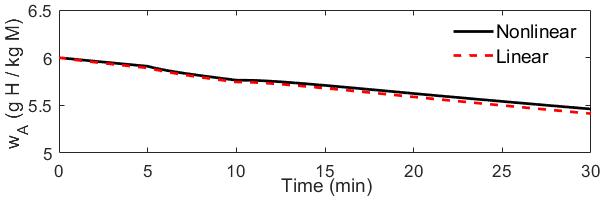}
         \caption{Hydrogen Weight Fraction in Reactor A}
         \label{fig:weightA_case2}
     \end{subfigure}
     \begin{subfigure}{0.49\textwidth}
         \centering
         \includegraphics[width=\textwidth]{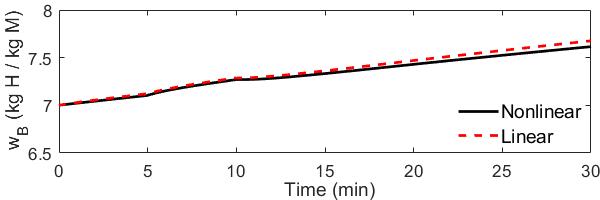}
         \caption{Hydrogen Weight Fraction in Reactor B}
         \label{fig:weightB_case2}
     \end{subfigure}
        \caption{Comparison of the state variables for Reactors A and B comparing the linear model (dashed lines) to the nonlinear model (solid lines) for Case 2 (hydrogen flowing from reactor B to reactor A).}
        \label{fig:Case2_lin_perf}
\end{figure}

\begin{figure}[!htb]
     \centering
     \begin{subfigure}{0.49\textwidth}
         \centering
         \includegraphics[width=\textwidth]{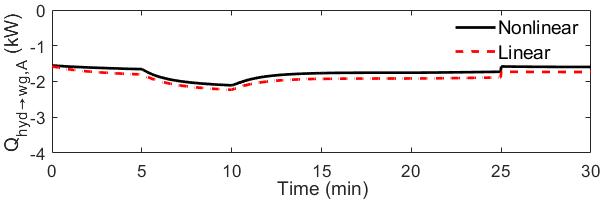}
         \caption{Reactor A}
         \label{fig:heatA_case2}
     \end{subfigure}
     \begin{subfigure}{0.49\textwidth}
         \centering
         \includegraphics[width=\textwidth]{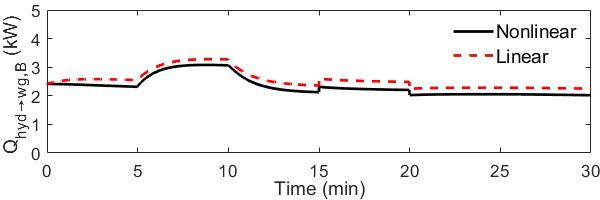}
         \caption{Reactor B}
         \label{fig:heatB_case2}
     \end{subfigure}
     \caption{Comparison of the heat transfer rates out of Reactors A and B comparing the linear model (dashed lines) to the nonlinear model (solid lines) for Case 2 (hydrogen flowing from reactor B to reactor A).}
        \label{fig:Case2_lin_perf_heat}
\end{figure}

 To determine the suitability of the linear state-space model for use in controller design, the dynamics predicted by the linear model are compared to those predicted by the nonlinear model for a given set of input conditions. To compare the models, the linear and nonlinear governing equations are simulated in MATLAB using a variable-step solver that uses fifth-order numerical differentiation. We consider two cases.  In Case 1, the initial conditions are defined so that hydrogen is desorbed in reactor B and flows to reactor A, where it is absorbed, as shown in Figures~\ref{fig:Case1_lin_perf} and~\ref{fig:Case1_lin_perf_heat}. In Case 2, the initial conditions are defined such that hydrogen is desorbed in reactor A and flows to reactor B where it is absorbed, as shown in Figures~\ref{fig:Case2_lin_perf} and~\ref{fig:Case2_lin_perf_heat}. These two cases, with opposite reactions occurring in the reactors, represent the charging and discharging modes in an energy storage system. Note that which mode is charging and which is discharging will depend on which reactor is being used to deliver the load. In both cases, each of the three control input variables is perturbed for a five-minute period, and the dynamic response of each model is observed. These perturbations to the input variables for both cases are shown in Figure~\ref{fig:Case1&2_inputs}. For both cases, the initial values of the state variables ($x_{0}$), as well as the input and disturbance variables ($u_{0}$), are listed in Table~\ref{tab:in_var}, and each simulation is initialized at the linearization point $(x_0,u_0)$.

\begin{table}[!htb]
\caption{Initial values for the state and input variables. Note that kg H indicates mass of hydrogen and kg M indicates mass of the metal hydride.} 
\begin{center}
  \begin{tabular}{l c c c}
  \noalign{\vskip -1.5mm}
  \hline
  \noalign{\vskip 1mm}
  $\textbf{\text{Variable}}$ & $\textbf{\text{Case 1}}$ & $\textbf{\text{Case 2}}$ & $\textbf{\text{Units}}$\\
  \noalign{\vskip 1mm}
  \hline
  \noalign{\vskip 1mm}
    $T_{hyd,A}$  & 6.89 & 6.89 & \degree C \\
    $T_{hyd,B}$ & 36.9 & 34.9 & \degree C \\
    $P_{H,A}$  & 480 & 290 & kPa \\ 
    $P_{H,B}$ & 290 & 360 & kPa \\
    $w_{A}$ & 0.006 & 0.007 & kg H / kg M \\
    $w_{B}$ & 0.006 & 0.007 & kg H / kg M \\
    $\dot{m}_{wg,A}$  & 0.2 & 0.2 & kg/s \\
    $\dot{m}_{wg,B}$ & 0.2 & 0.2 & kg/s \\
    $\Delta P_{comp}$  & 210 & 80 & kPa \\ 
    $T_{wg,in,A}$ & 1.89 & 11.9 & \degree C\\
    $T_{wg,in,B}$ & 42.9 & 30.9 & \degree C \\
  \noalign{\vskip 1mm}
  \hline
  \end{tabular}
\end{center}
\label{tab:in_var}
\end{table}

\par In both cases, the linear model matches the nonlinear model closely. This is quantified using the root mean square error (RMSE) between the state variables predicted by the linear model and the nonlinear model, as shown in Table \ref{tab:rmse_case1}. The normalized RMSE (NRMSE) values were calculated using Equation~\ref{eq:NRMSE}. All variables of the same type (e.g. temperature, pressure, or weight fraction) were normalized against the same value, determined by finding the maximum range over which each variable varies in either reactor (A or B) in either of the two simulation cases (see Equations~\ref{eq:delxA_NMRSE}-~\ref{eq:delxB_NMRSE}, where the subscript $i$ refers to the initial value of a given state variable simulated for a given case study).

\begin{equation}\label{eq:NRMSE}
NMRSE =\ \frac{RMSE}{max(\Delta x_{A},\Delta x_{B})}
\end{equation}

\begin{multline}\label{eq:delxA_NMRSE}
    \Delta x_{A} = max \left(max(x_{A} - x_{A,i})_{Case 1},max(x_{A} - x_{A,i})_{Case 2}\right) - \\ min \left(min(x_{A} - x_{A,i})_{Case 1},min(x_{A} - x_{A,i})_{Case 2}\right)
\end{multline}

\begin{multline}\label{eq:delxB_NMRSE}
    \Delta x_{B} = max \left(max(x_{B} - x_{B,i})_{Case 1},max(x_{B} - x_{B,i})_{Case 2}\right) - \\ min \left(min(x_{B} - x_{B,i})_{Case 1},min(x_{B} - x_{B,i})_{Case 2}\right)
\end{multline}

\begin{table}[!htb]
\caption{RMSE and NRMSE for all state variables for Case 1.} 
\begin{center}
  \resizebox{\hsize}{!}{\begin{tabular}{l c c c c c c}
  \noalign{\vskip -1.5mm}
  \hline
  \noalign{\vskip 1mm}
  & & & \textbf{RMSE} & & & \\
  $t (\text{min})$  & $T_{hyd,A} (\degree \text{C})$ & $T_{hyd,B} (\degree \text{C})$ & $P_{H,A} (\text{kPa})$ & $P_{H,B} (\text{kPa})$ & $w_{A} (\frac{\text{g H}}{\text{kg M}})$ & $w_{B} (\frac{\text{g H}}{\text{kg M}})$ \\
  \noalign{\vskip 1mm}
  \hline
  \noalign{\vskip 1mm}
    $0-5$  & 0.095 & 0.248 & 3.08 & 3.70 & 0.0047 & 0.0074 \\ 
    $5-10$ & 0.156 & 0.384 & 0.477 & 6.06 & 0.0060 & 0.017 \\ 
    $10-15$ & 0.094 & 0.354 & 2.66 & 7.23 & 0.0034 & 0.022 \\ 
    $15-20$ & 0.077 & 0.376 & 2.66 & 7.23 & 0.0061 & 0.029 \\
    $20-25$ & 0.057 & 0.388 & 0.847 & 6.70 & 0.0082 & 0.036 \\
    $25-30$ & 0.041 & 0.389 & 0.622 & 7.04 & 0.010 & 0.042 \\
  \noalign{\vskip 1mm}
  \hline
  \noalign{\vskip 1mm}
  & & & \textbf{NRMSE} & & & \\
  $t (\text{min})$  & $T_{hyd,A}$ & $T_{hyd,B}$ & $P_{H,A}$ & $P_{H,B}$ & $w_{A}$ & $w_{B}$ \\
  \noalign{\vskip 1mm}
  \hline
  \noalign{\vskip 1mm}
    $0-5$  & 2.42\% & 6.29\% & 2.26\% & 2.72\% & 0.356\% & 0.561\% \\ 
    $5-10$ & 3.96\% & 9.75\% & 0.349\% & 4.44\% & 0.455\% & 1.26\% \\ 
    $10-15$ & 2.37\% & 8.97\% & 1.95\% & 5.30\% & 0.258\% & 1.68\% \\ 
    $15-20$ & 1.94\% & 9.55\% & 0.918\% & 4.58\% & 0.462\% & 2.20\% \\ 
    $20-25$ & 1.45\% & 9.84\% & 0.621\% & 4.92\% & 0.621\% & 2.69\% \\ 
    $25-30$ & 1.05\% & 9.88\% & 0.456\% & 5.16\% & 0.780\% & 3.19\% \\ 
  \noalign{\vskip 1mm}
  \hline
  \end{tabular}}
\end{center}
\label{tab:rmse_case1}
\end{table}

\par For Case 1, the fastest response in the system is that of the hydrogen pressure (seen in Figures~\ref{fig:pressA_case1} and~\ref{fig:pressB_case1}) to the change in the pressure difference created by the compressor ($\Delta P_{comp}$) at $t=5\min$ and at $t=10\min$. At $t=5\min$, increasing $\Delta P_{comp}$ results in an almost immediate change in the hydrogen pressure of reactor A ($P_{H,A}$)  approximately equal to the change in $\Delta P_{comp}$, as seen in Figure \ref{fig:pressA_case1}. To understand why $P_{H,A}$ changes more than $P_{H,B}$, we can revisit the two terms of the pressure balance given in Equation~\ref{eq:pbal} (repeated below for convenience): the absorption or desorption rate (depending on the direction of the reaction) $rm_{hyd}$, and the hydrogen mass flow rate into the reactor, $\dot{m}_{H,in}$. 
\begin{equation}\nonumber
\frac{dP_{H}}{dt} =\ \frac{RT_{hyd}}{\phi V_{shell}}\left (rm_{hyd} + \dot{m}_{H,in}  \right )
\end{equation}
At this operating condition, $r_{B}$ (the reaction rate on a per unit mass of hydride basis) is more sensitive to changes in pressure than $r_{A}$. When $\Delta P_{comp}$ increases, the magnitude of $\dot{m}_{H,in}$ increases in each reactor. This term becomes much larger than the absorption rate in reactor A or the desorption rate in reactor B. However, the desorption rate in reactor B increases enough to balance the mass flow rate after only a small change in $P_{H,B}$, while it takes a much larger change in $P_{H,A}$ before the absorption rate in reactor A balances the mass flow rate.
Thus, $P_{H,A}$ increases much more than $P_{H,B}$ decreases. This process occurs very quickly; in the initial second after $\Delta P_{comp}$ increases, the driving pressure difference $P_{H,B}+\Delta P_{comp} - P_{H,A}$ decreases by 94.8\%.  Thus, it appears as a step change when plotted over a 30-minute time span in Figure \ref{fig:Case1_lin_perf}c. The linear state-space model successfully captures this response, as seen in Figures \ref{fig:pressA_case1} and \ref{fig:pressB_case1}.  
\par
\begin{figure}[!htb]
     \centering
     \begin{subfigure}{0.49\textwidth}
         \centering
         \includegraphics[width=\textwidth]{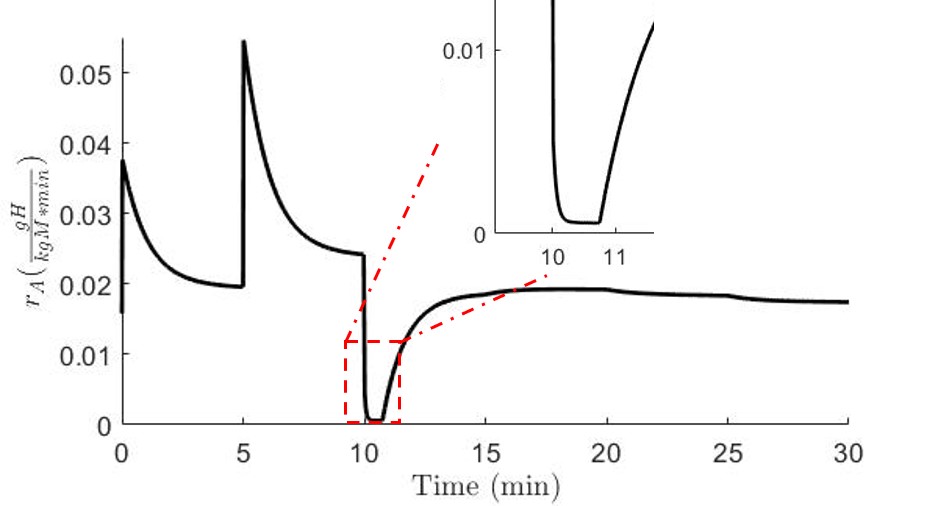}
         \caption{Reactor A}
         \label{fig:rrateA_case1}
     \end{subfigure}
     \begin{subfigure}{0.49\textwidth}
         \centering
         \includegraphics[width=\textwidth]{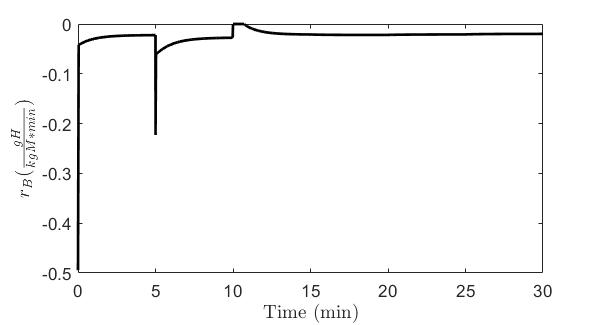}
         \caption{Reactor B}
         \label{fig:rrateB_case1}
     \end{subfigure}
     \caption{Reaction rates in Reactors A and B for Case 1 (hydrogen flowing from reactor A to reactor B).}
        \label{fig:Case1_rrate}
\end{figure}

 At $t=10\min$, when $\Delta P_{comp}$ returns to its original value, there is a significant step change in both pressures. The change in $\Delta P_{comp}$ causes hydrogen to flow from reactor A to reactor B, whereas it otherwise flows from reactor B to reactor A in Case 1.  Therefore, in both reactors, $rm_{hyd}$ and $\dot{m}_{H,in}$ have the same sign, so $r_{B}m_{hyd,B}$ cannot balance $\dot{m}_{H,in,B}$ as it did before. Once the initial pressure difference has been reduced, the mass flow is reduced to a small value, so the changes in pressure are primarily due to the reaction rates. However, the response of $r_{B}$ to the change in the pressure causes $r_{B}$ to go to zero, as shown in Figure~\ref{fig:rrateB_case1}. As shown in Figure~\ref{fig:rrateA_case1}, however, $r_{A}$ decreases but does not go to zero. This means that for approximately 45 seconds, hydrogen is being absorbed in one reactor but not desorbed in the other. Thus, the pressure in both reactors slowly decreases, since hydrogen is flowing out of reactor B, but the absorption rate in reactor A is larger than the flow rate in. After $\sim$45 seconds, the increasing temperature seen in Figure~\ref{fig:tempB_case1} causes the absorption rate for reactor B to grow larger than $\dot{m}_{H,in,B}$, so $P_{H,B}$ starts increasing. This increase leads to an increased mass flow rate to reactor A, which is larger than the absorption rate in reactor A, resulting in $P_{H,A}$ increasing as well.  

The most significant difference between the linear and nonlinear models can be seen in Figure~\ref{fig:pressB_case1}, where the linear model does not capture the initial spike in $P_{H,B}$ and the decay that follows it, but accurately captures the behavior of the pressure after that time. This is because the reaction rate equation (Equation~\ref{eq:rrate}) is discontinuous when it goes to zero; as shown in Equation~\ref{eq:rrate}. Since the linear model does not include this discontinuity, it fails to capture the behavior of the pressure while the reaction rate is equal to zero. However, once the reaction rate is again nonzero, the linear model again follows the nonlinear model. 
\par The response of the hydride temperature in each reactor to the change in $\Delta P_{comp}$ is shown in Figures \ref{fig:tempA_case1} and \ref{fig:tempB_case1}. Here, the increased reaction rate after the increase in $\Delta P_{comp}$ results in an increase in the heat released or absorbed by the reaction, pushing the temperature of the hydrides in both reactors away from the temperature of the circulating fluid in their reactors. This causes the heat transfer rate between the hydride in each reactor and the circulating fluid to increase, as shown in Figure \ref{fig:Case1_lin_perf_heat}. Once these heat transfer rates are approximately equal to the heat transfer from  absorption and desorption, the temperatures become close to steady. While the linear model underestimates $T_{hyd,B}$ after the step change, as shown in Figure \ref{fig:tempB_case1} at $t=5$ min, it stays within 10\% of the nonlinear model, and $T_{hyd,A}$ stays within 5\% throughout the simulation. 
\par In contrast to the visible changes in pressure and temperature that occur when $\Delta P_{comp}$ changes, there is not a significant change in these variables when either of the mass flow rates are changed. However, as shown in Figure~\ref{fig:Case1_lin_perf_heat}, there is a step change in the heat transfer rate in each reactor when there is a change in the mass flow rate in that reactor. This is because the heat transfer rate is a linear function of the mass flow rate. This change is accurately captured by the linear model, as seen in Figures~\ref{fig:heatA_case1} and \ref{fig:heatB_case1}. All variables stay within 10\% of the nonlinear model throughout the simulation.
\par Finally, the change in the weight fraction over time is different from that of the hydride temperature and pressure because the weight fraction changes continually (except for $w_{B}$ while $r_{B}$ is equal to zero) and does not enter a near-equilibrium state. The dynamics of the weight fraction are controlled by the reaction rate, which is the derivative of the weight fraction. As discussed in regards to the pressure dynamics, the reaction rate in both reactors changes quickly in the seconds after the change in $\Delta P_{comp}$ until $r_{A}m_{hyd,A}$, $r_{B}m_{hyd,B}$, and $\dot{m}_{H,in}$ have approximately equal magnitudes. This quick change in the reaction rate, combined with the very slow change for the rest of the simulation period, results in the weight fraction in each reactor resembling a linear function with a different slope after $t=5\min$, as seen in Figures \ref{fig:weightA_case1} and \ref{fig:weightB_case1}. After $\Delta P_{comp}$ returns to its original value at $t=10\min$, the rate of change of the weight fraction also returns to approximately its original value. For the weight fraction, the error of the linear models stays below 4\% for both reactors as shown in Table \ref{tab:rmse_case1}. Like the pressure, the weight fraction does not respond significantly to changes in the mass flow rates.

\begin{table}[!htb]
\caption{RMSE and NRMSE for all state variables for Case 2.} 
\begin{center}
  \resizebox{\hsize}{!}{\begin{tabular}{l c c c c c c}
  \noalign{\vskip -1.5mm}
  \hline
  \noalign{\vskip 1mm}
    & & & \textbf{RMSE} & & & \\
  $t (\text{min})$  & $T_{hyd,A} (\degree \text{C})$ & $T_{hyd,B} (\degree \text{C})$ & $P_{H,A} (\text{kPa})$ & $P_{H,B} (\text{kPa})$ & $w_{A} (\frac{\text{g H}}{\text{kg M}})$ & $w_{B} (\frac{\text{g H}}{\text{kg M}})$ \\
  \noalign{\vskip 1mm}
  \hline
  \noalign{\vskip 1mm}
    $0-5$  & 0.273 & 0.380 & 1.01 & 5.22 & 0.011 & 0.012 \\
    $5-10$ & 0.317 & 0.400 & 0.829 & 5.55 & 0.017 & 0.017 \\
    $10-15$ & 0.391 & 0.455 & 1.16 & 7.46 & 0.025 & 0.023 \\ 
    $15-20$ & 0.428 & 0.474 & 0.787 & 7.01 & 0.033 & 0.034 \\
    $20-25$ & 0.424 & 0.476 & 0.980 & 7.29 & 0.039 & 0.045 \\
    $25-30$ & 0.407 & 0.479 & 1.17 & 7.58 & 0.045 & 0.056 \\
  \noalign{\vskip 1mm}
  \hline
    & & & \textbf{NRMSE} & & & \\
  $t (\text{min})$  & $T_{hyd,A}$ & $T_{hyd,B}$ & $P_{H,A}$ & $P_{H,B}$ & $w_{A}$ & $w_{B}$ \\
  \noalign{\vskip 1mm}
  \hline
  \noalign{\vskip 1mm}
    $0-5$  & 6.92\% & 9.66\% & 0.743\% & 3.83\% & 0.811\% & 0.886\% \\ 
    $5-10$ & 8.05\% & 10.2\% & 0.608\% & 4.07\% & 1.30\% & 1.27\% \\ 
    $10-15$ & 9.93\% & 11.6\% & 0.851\% & 5.47\% & 1.89\% & 1.75\% \\ 
    $15-20$ & 10.9\% & 12.0\% & 0.577\% & 5.14\% & 2.46\% & 2.61\% \\ 
    $20-25$ & 10.8\% & 12.1\% & 0.718\% & 5.34\% & 2.95\% & 3.43\% \\ 
    $25-30$ & 10.3\% & 12.1\% & 0.858\% & 5.55\% & 3.42\% & 4.23\% \\ 
  \noalign{\vskip 1mm}
  \hline
  \end{tabular}}
\end{center}
\label{tab:rmse_case2}
\end{table}
 
\par For Case 2, despite hydrogen flow in the system moving in the opposite direction, the same trends can be seen as discussed for Case 1. The pressure dynamics are similar, including $r_{B}$ going to zero for some time after $t=10$ min. However, because reactor A is desorbing hydrogen here rather than absorbing it as in Case 1, the pressure in both reactors increases while $r_{B}$ equals zero, and then decreases once absorption starts in reactor B, as seen in Figures~\ref{fig:pressA_case2} and~\ref{fig:pressB_case2}. The same general trends can also be seen in the temperature and weight fraction dynamics, with the only differences being due to the reversal of which reactor is absorbing hydrogen and releasing heat to the circulating fluid, and which is desorbing hydrogen and being heated by the circulating fluid.


\par Overall, the linear model predictions match those of the nonlinear model well in Case 2. As seen in Table~\ref{tab:rmse_case2}, there is a larger error for the hydride temperature states in this case than for Case 1, but the linear model is still accurate within 12.5\% across the entire simulation period. The error for $P_{H,A}$ is lower for this case, remaining within 1\% of the nonlinear model, while the highest error for $P_{H,B}$ is still less than 6\% different from the nonlinear model. The error between the linear and nonlinear model predictions for the weight fraction states follows a similar pattern to Case 1, continually increasing over time while never exceeding 5\%, without any of the changes to the inputs noticeably affecting the rate at which the error increases.

\subsection{Resetting the Linearization Point}

\par
\begin{figure}[!htb]
     \centering
     \begin{subfigure}{0.49\textwidth}
         \centering
         \includegraphics[width=\textwidth]{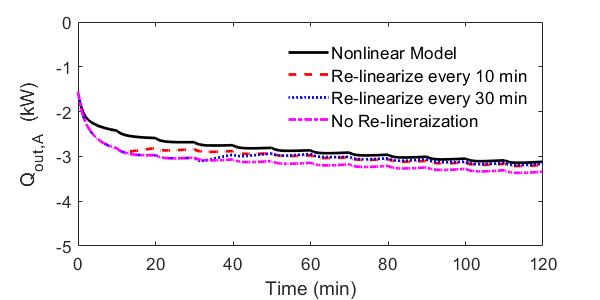}
         \caption{Reactor A}
         \label{fig:heatA_relin}
     \end{subfigure}
     \begin{subfigure}{0.49\textwidth}
         \centering
         \includegraphics[width=\textwidth]{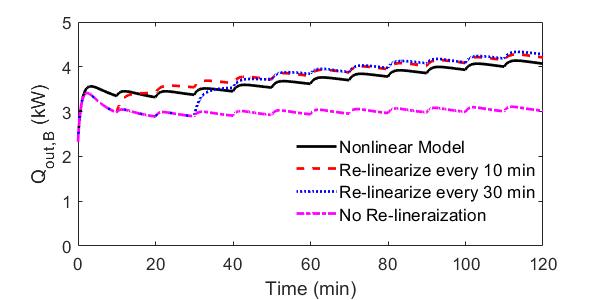}
         \caption{Reactor B}
         \label{fig:heatB_relin}
     \end{subfigure}
     \caption{Heat transfer rates from the hydride beds to the circulating fluid in each reactor. The performance of the linear models is compared to the nonlinear model (solid black line) for three different cases: one where the model is never re-linearized (dot-dashed pink line), one where it is re-linearized every 10 minutes (dotted blue line), and one where it is re-linearized every 30 minutes (dashed red line). Throughout this study, the compressor pressure difference is increased by 10 kPa every 10 minutes.}
    \label{fig:Qout_relin_with_no_relin}
\end{figure}

 While the validation results show that the linear model predicts the system dynamics accurately within approximately 12.5\% of the linearization point, it is expected that during operation, the system will deviate further from a single point.  This can adversely affect the controller design if it assumes the dynamics do not vary from a single linear model. One way to address this is to periodically re-linearize the model around the current operating condition.  To illustrate how re-linearization affects the accuracy of the linear model, we simulate both the nonlinear and linear models for 120 minutes, beginning with the initial conditions shown in Table~\ref{tab:in_var_relin}. In order to sustain the absorption and desorption reactions in the reactors for 120 minutes, the compressor pressure difference is increased by 10 kPa every 10 minutes. We compare three different cases: 1) simulating the linear model from the same initial conditions without any re-linearization, 2) simulating the linear model beginning with the same initial conditions but then re-linearizing it every 10 minutes, and 3) simulating the linear model beginning with the same initial conditions but then re-linearizing it every 30 minutes. The resulting heat transfer rates for these cases are compared against the nonlinear model simulation in Figure~\ref{fig:Qout_relin_with_no_relin}, and the RMSE for the heat transfer rates is given in Table~\ref{tab:rmse_relin}. It is worth noting that because the linearized model is fully parameterized, re-linearizing is equivalent to updating the model parameters.  In other words, re-linearizing does not contribute any significant computational complexity to the numerical simulation.

\begin{table}[!htb]
\caption{Initial values for the state and input variables for comparing performance with and without re-linearizing the model and for the case in which the controller is used for flow from reactor B to reactor A.} 
\begin{center}
  \begin{tabular}{l c c l c c}
  \noalign{\vskip -1.5mm}
  \hline
  \noalign{\vskip 1mm}
  $\textbf{\text{Variable}}$ & $\textbf{\text{Value}}$  & $\textbf{\text{Units}}$ & $\textbf{\text{Variable}}$ & $\textbf{\text{Value}}$  & $\textbf{\text{Units}}$\\
  \noalign{\vskip 1mm}
  \hline
  \noalign{\vskip 1mm}
    $T_{i,A}$ & 6.9 & \degree C & $\dot{m}_{wg,A}$  & 0.2 & kg/s \\
    $T_{i,B}$ & 34. & \degree C & $\dot{m}_{wg,B}$  & 0.2 & kg/s \\
    $P_{i,A}$ & 290 & kPa & $\Delta P_{comp}$  & 80 & kPa \\ 
    $P_{i,B}$ & 360 & kPa & $T_{wg,in,A}$ & 12 & \degree C\\
    $w_{i,A}$ & 0.008 & kg H / kg M & $T_{wg,in,B}$ & 27 & \degree C \\
    $w_{i,B}$ & 0.0045 & kg H / kg M & & & \\
  \noalign{\vskip 1mm}
  \hline
  \end{tabular}
\end{center}
\label{tab:in_var_relin}
\end{table}

\begin{table}[!htb]
\caption{RMSE for the heat rates for the 3 cases studied. All values are in kW.} 
\begin{center}
  \begin{tabular}{l c c c c c c}
  \noalign{\vskip -1.5mm}
  \hline
  \noalign{\vskip 1mm}
    & \multicolumn{2}{c}{No Relin.} & \multicolumn{2}{c}{Every 10 min} & \multicolumn{2}{c}{Every 30 min} \\
  $t (\text{min})$  & $Q_{A}$ & $Q_{B}$ & $Q_{A}$ & $Q_{B}$ & $Q_{A}$ & $Q_{B}$ \\
  \noalign{\vskip 1mm}
  \hline
  \noalign{\vskip 1mm}
    $0-30$ & 0.338 & 0.376 & 0.251 & 0.154 & 0.338 & 0.376 \\
    $30-60$ & 0.308 & 0.596 & 0.127 & 0.179 & 0.186 & 0.148 \\
    $60-90$ & 0.257 & 0.796 & 0.101 & 0.166 & 0.081 & 0.196 \\
    $90-120$ & 0.228 & 0.983 & 0.081 & 0.145 & 0.061 & 0.182 \\
  \noalign{\vskip 1mm}
  \hline
  \end{tabular}
\end{center}
\label{tab:rmse_relin}
\end{table}

\par From Figure \ref{fig:Qout_relin_with_no_relin} we can see that re-linearization does improve the accuracy of the linear model over the duration of the simulation. For both heat transfer rates, both cases with re-linearization match the nonlinear model after re-linearization more closely than the case without re-linearization. It is worth noting that the extent of the nonlinearity of each reactor is not the same. For reactor A, as shown in Figure~\ref{fig:heatA_relin}, even without re-linearization, the linear model matches the nonlinear one on average within 0.35 kW. Furthermore, the linear model does not diverge from the nonlinear model: the error for the last 30 minutes is less than that for the first 30 minutes. For reactor B, as shown in Figure \ref{fig:heatB_relin}, the model with no re-linearization diverges from the nonlinear model almost immediately, with error increasing over time. 
\par Comparing the performance of the linear model with re-linearization every 10 minutes to that with re-linearization every 30 minutes, we can see in Table~\ref{tab:rmse_relin} that both stay within 0.2 kW of the nonlinear model after the first 30 minutes. The difference in error between them is small, with more frequent re-linearization never resulting in a reduction in error of more than 0.03 kW once both models have been re-linearized at least once. In general, the frequency with which the parameters of the linear model should be updated will depend on how close the last linearization point is to the ``near-equilibrium'' state in which there is only a slow change in the temperature, pressure, and reaction rate in each reactor. For the operating conditions considered here, we can conclude that re-linearizing every 30 minutes is sufficient for maintaining accuracy; this will be applied to the controller design discussed in the next section. 

%% file: 03a_Control_Design.tex
\section{Controller Design and Synthesis}
\label{sec:control}

In this section, we describe the design of a multi-input-multi-output (MIMO) model predictive controller (MPC) for regulating the operation of the proposed two-reactor metal hydride thermal energy storage system. 

\subsection{Model Predictive Control Algorithm}
MPC is a control technique that utilizes a model of the system dynamics to optimize a sequence of control inputs over a specified time horizon by predicting the dynamical response of the system over said horizon based on different sequences of inputs \cite{Camacho2013ModelControl}. At each measurement sample, MPC involves solving an $N$-step ahead online optimization problem to predict the optimal sequence of control inputs $\left[u\left(k\right),\ u\left(k+1\right),\cdots u(k+N)\right]$ that will drive the output sequence $\left[y\left(k\right),\ y\left(k+1\right),\cdots y(k+N)\right]$ toward a desired reference trajectory. At each time in the control sampling period, the MPC problem is solved, and the optimized variables $u\left(k\right)$ are taken as the control input at that sample instant. 

Given the nonlinear dynamics of the hydride system, a successive linearization technique \cite{Henson1998NonlinearDirections} is used with a linear MPC design to achieve the desired control objectives.  As was shown in Sec.~\ref{sec:modelval}, local linearization of the model matches the nonlinear model within 12.1\% (based upon the validation presented in Section \ref{sec:modelval}) and is therefore adequate for use as a prediction model in a MPC design. We design and implement the MPC as a discrete-time controller.  Therefore, the state dynamics and system outputs are discretized and expressed as 

\begin{align}
\label{eq:discretemodel}
x\left(k+1\right)&=\mathbf{A}x\left(k\right)+\mathbf{B}u
\left(k\right)+\mathbf{B}_{d}d(k)  \\
{y}\left(k\right)&=\mathbf{C}x\left(k\right)+\mathbf{D}u\left(k\right)+\mathbf{D}_{d}d(k) \enspace .	\nonumber  
\end{align}

In these equations, the input vector has been separated into $u$ and $d$, where $u$ is the set of input variables we can control (mass flow rates and compressor pressure difference), while $d$ is the set of disturbance inputs (circulating fluid inlet temperatures), i.e. those input variables that cannot be controlled . To achieve zero steady-state tracking error, we augment the linear-quadratic regulator (LQR) with integral control as shown in Equation~\eqref{eq:augssmodel2}:

\begin{align}
\label{eq:augssmodel2}
\tilde{x}\left(k+1\right)&=\tilde{\mathbf{A}}\tilde{x}\left(k\right)+\tilde{\mathbf{B}}u\left(k\right)+{\tilde{\mathbf{B}}}_\mathbf{d}\tilde{d}(k)    \\
\tilde{y}\left(k\right)&=\tilde{\mathbf{C}}\tilde{x}(k) \enspace , \nonumber
\end{align}

\noindent where
\begin{equation}
    \tilde{\mathbf{A}}=\left[\begin{matrix}\mathbf{A}&\mathbf{0}\\\mathbf{C}&\mathbf{0}\\\end{matrix}\right],   \tilde{\mathbf{B}}=\left[\begin{matrix}\mathbf{B}\\\mathbf{D}\\\end{matrix}\right],\ {\tilde{\mathbf{B}}}_\mathbf{d}=\left[\begin{matrix}\mathbf{B}_\mathbf{d}&\mathbf{0}\\\mathbf{D}_\mathbf{d}&-\mathbf{I}\\\end{matrix}\right],\ \tilde{\mathbf{C}}=\left[\begin{matrix}\mathbf{0}&\mathbf{I}\\\end{matrix}\right],
\label{eq:augssmodelmatrices}
\end{equation}

\noindent and $ \tilde{x}\left(k\right)=\left[\begin{matrix}x\left(k\right),&x_i\left(k\right)\\\end{matrix}\right]^T$ and $\tilde{d}(k)=\left[\begin{matrix}d\left(k\right)&r\left(k\right)\\\end{matrix}\right]^T$. Based on Equation \eqref{eq:augssmodelmatrices}, the output of the control model simplifies to $\tilde{y}\left(k\right)=x_i\left(k\right)$ where $x_i\left(k+1\right)=\dot{Q}\left(k\right)-r(k)$. Therefore, the MPC is formulated as an error regulation problem to drive $x_i(k)$ to zero, which is equivalent to driving the heat transfer rates $\dot{Q}\left(k\right)$ to the reference values $r(k)$. The MPC problem can be stated as

{\begin{equation}
\begin{aligned}
\min_{\mathbf{U},\mathbf{Y}} \quad & J = \sum_{k=1}^{N} \widetilde{y}\left(k\right)^T\mathbf{Q}\widetilde{y}\left(k\right) + u\left(k\right)^T\mathbf{R}u\left(k\right)  \\
\textrm{s.t.} \quad & \widetilde{x}\left(k+1\right) = \widetilde{\mathbf{A}}\widetilde{x}\left(k\right) + \widetilde{\mathbf{B}}u\left(k\right) + \mathbf{\widetilde{B}_d}\widetilde{d}\left(k\right) \enspace \forall \enspace k \in [1,N] \\ 
& \widetilde{y}\left(k+1\right) = \widetilde{\mathbf{C}}\widetilde{x}\left(k+1\right)  \enspace \forall \enspace k \in [1,N] \\
& u_{min} \leq u\left(k\right) \leq u_{max} \enspace \forall \enspace k \in [1,N] \\
& |u\left(k\right) - u\left(k+1\right)| \leq \delta u_{max} \enspace \forall \enspace k \in [1,N-1] \enspace ,
\end{aligned}
\label{eq:mpcproblem}
\end{equation}}
\smallskip

\noindent 

\noindent where

{\small
\begin{equation}
\begin{aligned}
& \mathbf{U} = [u\left(k=1\right), \enspace u\left(2\right), \enspace \hdots \enspace, \enspace u\left(N\right)] \\
& \mathbf{Y} = [\widetilde{y}\left(k=1\right), \enspace \widetilde{y}\left(2\right), \enspace \hdots \enspace, \enspace \widetilde{y}\left(N\right)] \enspace ,
\end{aligned} \nonumber 
\end{equation}}
\smallskip

\noindent and $\mathbf{Q}$ and $\mathbf{R}$ are positive definite weighting matrices. The constraints ensure (1) adherence to the state dynamics prescribed by the control model given in Equations~\eqref{eq:augssmodel2} and \eqref{eq:augssmodelmatrices}, (2) that the optimal values for the inputs $u$ are within the achievable range of inputs, and (3) that the rate of change in the chosen inputs does not exceed actuator limits. 

\subsection{Controller Synthesis}
\par For the simulated case studies, the results of which are presented in Section \ref{sec:results}, the matrices $\mathbf{Q}$ and $\mathbf{R}$ are tuned to penalize tracking error and the magnitude of the control inputs, respectively. The relative magnitude of the weights in these matrices drives the controller performance, and can be motivated by the dynamics of the system itself. One representative linearization of the model, corresponding to Case 1 as described later in Section \ref{sec:results}, is shown in Equations~\eqref{eq:case1_f}~-~\eqref{eq:case1_D}.

In the two-reactor hydride system, the output variables $\dot{Q}_{hyd \rightarrow wg,m}$ and $\dot{Q}_{hyd \rightarrow wg,n}$ are directly affected by the mass flow rate of the circulating fluid in each reactor, $\dot{m}_{wg,m}$ and $\dot{m}_{wg,n}$, respectively, through the $\mathbf{D}$ matrix. Note that these correspond to the first two inputs defined in the input vector $u$.  The third input variable, the compressor pressure difference $\Delta P_{comp}$, does not directly affect the output variables; it only indirectly affects them through its effect on the hydride temperature states, $T_{hyd,m}$ and $T_{hyd,n}$.  Therefore, in tuning $\mathbf{Q}$ and $\mathbf{R}$, we more heavily penalize $\Delta P_{comp}$, relative to the weights placed on $\dot{m}_{wg,m}$ and $\dot{m}_{wg,n}$, to incentivize the controller to use $\Delta P_{comp}$ to help regulate  $\dot{Q}_{hyd \rightarrow wg,m}$ and $\dot{Q}_{hyd \rightarrow wg,n}$.  The final weights chosen for $\mathbf{Q}$ and $\mathbf{R}$ are

\begin{equation}\label{eq:Qvals}
\mathbf{Q}=\ \begin{bmatrix}100
& 0 \\ 0 & 100
\end{bmatrix}
\end{equation}

\begin{equation}\label{eq:Rvals}
\mathbf{R}=\ \begin{bmatrix}3\times10^{11}
& 0 & 0 \\ 0 & 3\times10^{11} & 0 \\ 0 & 0 & 1 
\end{bmatrix}\enspace .
\end{equation}

We formulate and solve the MPC as a quadratic program using the \emph{quadprog} solver within the MATLAB Optimization Toolbox \cite{TheMathWorksInc.2018MatlabToolbox}. Formulating the problem as a quadratic program permits a computationally-efficient controller synthesis to be obtained for each control sample with little computational overhead.



\begin{equation}
\label{eq:case1_f}
f(x,u) = \frac{d}{dt}\left( \begin{bmatrix} T_{hyd,m} \\ P_{H,m} \\ w_{m} \\ T_{hyd,n} \\ P_{H,n} \\ w_{hyd,n} \end{bmatrix} \right) = \mathbf{A} \begin{bmatrix}T_{hyd,m} \\ P_{H,m} \\ w_{m} \\ T_{hyd,n} \\ P_{H,n} \\ w_{hyd,n}
\end{bmatrix} + \mathbf{B} \begin{bmatrix} \dot{m}_{wg,m} \\ \dot{m}_{wg,n} \\ \Delta P_{comp} \\ T_{wg,in,m} \\ T_{wg,in,n}\end{bmatrix}
\end{equation}

\begin{equation}
\label{eq:case1_A}
\resizebox{0.9\hsize}{!}{$\mathbf{A} = \begin{bmatrix} -8.17\times 10^{-3} & -3.35\times10^{-8} & -8.13 & -9.07\times10^{-5} & 4.09\times10^{-7} & 0 \\ 8840 & -4.38 & 1.11\times10^{7} & 0 & 3.96 & 0 \\ -2.28\times10^{-7} & 1.38\times10^{-11} & -2.93\times10^{-4} & 0 & 0 & 0 \\ 0 & 0 & 0 & -0.216 & 1.83\times10^{-5} & -94.4 \\ 0 & 4.93 & 0 & 2.03\times10^{5} & -22.4 & 9.00\times10^{7} \\ 0 & 0 & 0 & -4.78\times10^{-6} & 4.09\times10^{-10} & -2.12\times10^{-3} \\ \end{bmatrix}$}
\end{equation}

\begin{equation}
\label{eq:case1_B}
\mathbf{B} = \begin{bmatrix} -0.039 & 0 & 4.47\times10^{-7} & 1.68\times10^{-3} & 0 \\ 0 & 0 & 3.91 & 0 & 0 \\ 0 & 0 & 0 & 0 & 0 \\ 0 & 0.064 & 0 & 0 & 2.99\times10^{-3} \\ 0 & 0 & -5.00 & 0 & 0 \\ 0 & 0 & 0 & 0 & 0 \\
\end{bmatrix}
\end{equation}

\begin{equation}
\label{eq:case1_g}
g(x,u) = \begin{bmatrix} \dot{Q}_{hyd \rightarrow wg,m} \\ \dot{Q}_{hyd \rightarrow wg,n}  \end{bmatrix}  = \mathbf{C} \begin{bmatrix}T_{hyd,m} \\ P_{H,m} \\ w_{m} \\ T_{hyd,n} \\ P_{H,n} \\ w_{hyd,n}
\end{bmatrix} + \mathbf{D} \begin{bmatrix} \dot{m}_{wg,m} \\ \dot{m}_{wg,n} \\ \Delta P_{comp} \\ T_{wg,in,m} \\ T_{wg,in,n}\end{bmatrix}
\end{equation}

\begin{equation}
\label{eq:case1_C}
\mathbf{C} = \begin{bmatrix} 367.1 & 0 & 0 & 0 & 0 & 0 \\ 0 & 0 & 0 & 436.3 & 0 & 0 \\ 
\end{bmatrix}
\end{equation}

\begin{equation}
\label{eq:case1_D}
\mathbf{D} = \begin{bmatrix} 7930 & 0 & 0 & -341 & 0 \\ 0 & -9600 & 0 & 0 & -452 \\
\end{bmatrix}
\end{equation}

%% file: 03b_Controller_Validation.tex
\section{Results}
\label{sec:results}
In this section, we implement the controller in simulation and demonstrate its performance in regulating the dynamics of the two-reactor metal hydride system through a series of case studies.

\subsection{Reference Tracking}
\par The MPC is designed primarily for referencing tracking, so we first verify its performance in the context of tracking variable heat transfer rates in each reactor. The model is simulated using the same initial conditions as those given in Table~\ref{tab:in_var}. Moreover, the control input variables are bounded based on the values shown in Table~\ref{tab:input_minmax}.

\begin{table}[!htb]
\caption{Upper and lower bounds on the control input variables.} 
\begin{center}
  \begin{tabular}{l c c c}
  \noalign{\vskip -1.5mm}
  \hline
  \noalign{\vskip 1mm}
  $\textbf{\text{Variable}}$ & $\textbf{\text{Units}}$ & $\textbf{\text{Minimum}}$ & $\textbf{\text{Maximum}}$ \\
  \noalign{\vskip 1mm}
  \hline
  \noalign{\vskip 1mm}
    $\dot{m}_{A}$ & kg/s & 0 & 0.8 \\
    $\dot{m}_{B}$ & kg/s & 0 & 0.8 \\
    $\Delta P_{comp}$ & kPa & 0 & 500 \\ 
  \noalign{\vskip 1mm}
  \hline
  \end{tabular}
\end{center}
\label{tab:input_minmax}
\end{table}

\par
\begin{figure}[!htb]
     \centering
     \begin{subfigure}{0.49\textwidth}
         \centering
         \includegraphics[width=\textwidth]{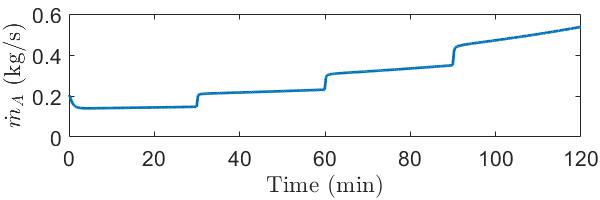}
         \caption{Circulating Fluid Mass Flow Rate A}
         \label{fig:mdotA_cntr_heat_case1}
         \vspace*{2.5mm}
     \end{subfigure}
     \begin{subfigure}{0.49\textwidth}
         \centering
         \includegraphics[width=\textwidth]{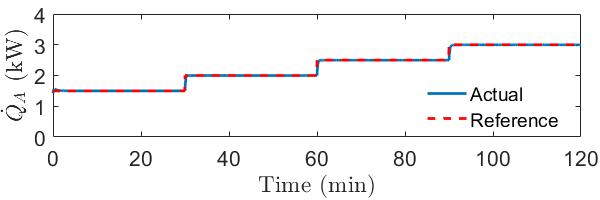}
         \caption{Heat Transfer Rate in Reactor A}
         \label{fig:heatA_cntr_heat_case1}
         \vspace*{2.5mm}
     \end{subfigure}
     \begin{subfigure}{0.49\textwidth}
         \centering
         \includegraphics[width=\textwidth]{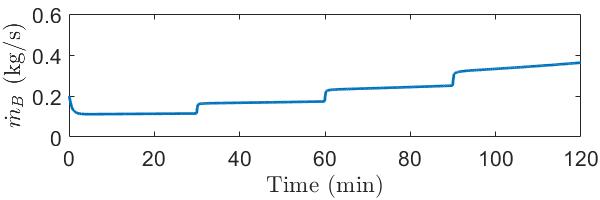}
         \caption{Circulating Fluid Mass Flow Rate B}
         \label{fig:mdotB_cntr_heat_case1}
         \vspace*{2.5mm}
     \end{subfigure}
     \begin{subfigure}{0.49\textwidth}
         \centering
         \includegraphics[width=\textwidth]{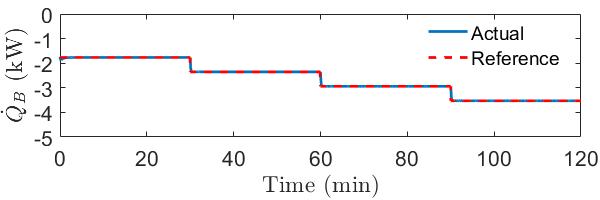}
         \caption{Heat Transfer Rate in Reactor B}
         \label{fig:heatB_cntr_heat_case1}
         \vspace*{2.5mm}
     \end{subfigure}
     \begin{subfigure}{0.49\textwidth}
         \centering
         \includegraphics[width=\textwidth]{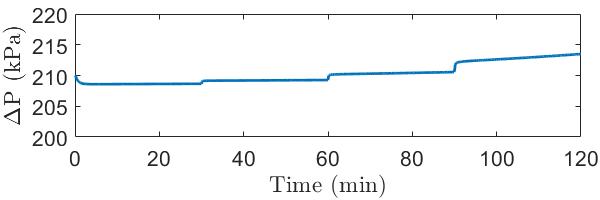}
         \caption{Compressor Pressure Difference}
         \label{fig:dP_cntr_heat_case1}
     \end{subfigure}
     \caption{Input values set by the controller and output heat transfer rates in the hydride reactors compared to the reference value for the case where hydrogen is being moved from reactor B to reactor A. The model is re-linearized and the reference value changed every 30 minutes.}
    \label{fig:cntr_charge_changeQ}
\end{figure}

 The model is simulated for a two-hour period with the desired heat transfer rate setpoints (reference values) changing every 30 minutes and the controller re-linearizing around the current operating point when the setpoint changes. The control input variables are updated at a frequency of 1 Hz. As with the model validation, two cases are considered, representing the two operating modes of the system.  In Case 1, the reference value for the heat transfer rate in Reactor A increases over a series of three step changes which are mirrored by step \emph{decreases} in the heat transfer rate in Reactor B.  The opposite trends in the references are demonstrated in Case 2.  The closed-loop simulation results for Case 1 are shown in Figure~\ref{fig:cntr_charge_changeQ}. The results for Case 2 are given in~\ref{sec:appA}.

\par Figure~\ref{fig:cntr_charge_changeQ} shows that the controller successfully tracks the step changes in the heat transfer rates in each reactor, doing so primarily through adjustments in each of the circulating fluid mass flow rates (see Figures~\ref{fig:mdotA_cntr_heat_case1} and~\ref{fig:mdotB_cntr_heat_case1}). This is consistent with the algebraic relationship between mass flow rate and heat transfer rate in each reactor. In addition, the mass flow rates slowly increase over time in between step changes, adjusting to maintain a fixed heat transfer rate while the temperature differential between each reactor and the associated circulating fluid decreases. As shown in Figure~\ref{fig:dP_cntr_heat_case1}, the controller makes less use of the compressor pressure difference to achieve its objectives. While its value does change over time, the difference between the minimum and maximum values of $\Delta P_{comp}$ is only around 1\% of the range of values the controller can set it to, while the difference between the minimum and maximum values of $\dot{m}_{A}$ is approximately 50\% of its range.

\par
\begin{figure}[!htb]
     \centering
     \begin{subfigure}{0.49\textwidth}
         \centering
         \includegraphics[width=\textwidth]{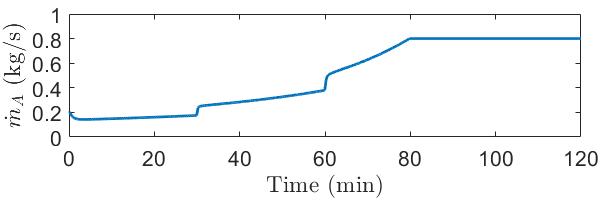}
         \caption{Circulating Fluid Mass Flow Rate A}
         \label{fig:mdotA_cntr_heat_case3}
         \vspace*{2.5mm}
     \end{subfigure}
     \begin{subfigure}{0.49\textwidth}
         \centering
         \includegraphics[width=\textwidth]{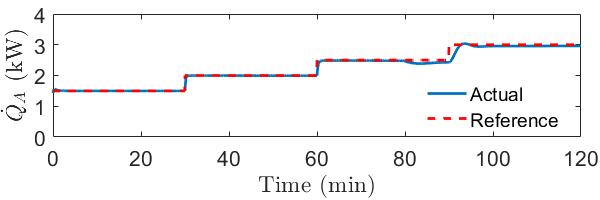}
         \caption{Heat Transfer Rate in Reactor A}
         \label{fig:heatA_cntr_heat_case3}
         \vspace*{2.5mm}
     \end{subfigure}
     \begin{subfigure}{0.49\textwidth}
         \centering
         \includegraphics[width=\textwidth]{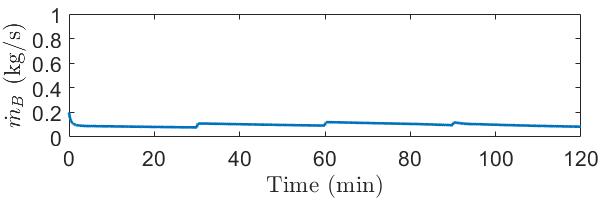}
         \caption{Circulating Fluid Mass Flow Rate B}
         \label{fig:mdotB_cntr_heat_case3}
         \vspace*{2.5mm}
     \end{subfigure}
     \begin{subfigure}{0.49\textwidth}
         \centering
         \includegraphics[width=\textwidth]{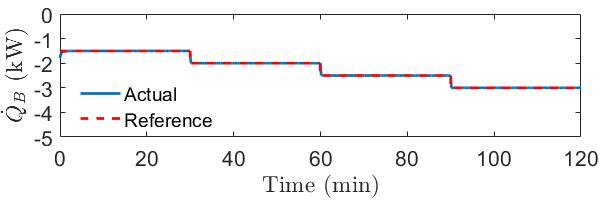}
         \caption{Heat Transfer Rate in Reactor B}
         \label{fig:heatB_cntr_heat_case3}
         \vspace*{2.5mm}
     \end{subfigure}
     \begin{subfigure}{0.49\textwidth}
         \centering
         \includegraphics[width=\textwidth]{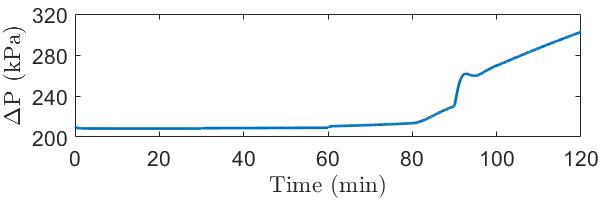}
         \caption{Compressor Pressure Difference}
         \label{fig:dP_cntr_heat_case3}
     \end{subfigure}
     \caption{Input values set by the controller and output heat transfer rates in the hydride reactors compared to the reference value for the case where hydrogen is being moved from reactor B to reactor A and the ratio between heat transfer rates is set to 1 instead of 0.85. The model is re-linearized and the reference value changed every 30 minutes.}
    \label{fig:cntr_charge_changeQ_1ratio}
\end{figure}

 For both this case and the case given in~\ref{sec:appA}, the ratio of the heat transfer rates in the two reactors are held constant even when the magnitude of the reference (desired) heat transfer rate changes. Specifically, the heat transfer rate in reactor A has the opposite sign and 85\% of the magnitude of that in reactor B. To see how the controller performance changes for a different ratio between the heat transfer rates, we consider a case with the same initial conditions and reference values for $\dot{Q}_{B}$ as defined in Case 1, but with the ratio of the magnitudes of the heat transfer rates set to 1. The control input signals and resulting heat transfer rates, compared to the reference values, are shown in Figure~\ref{fig:cntr_charge_changeQ_1ratio}.

\par A major departure in these results, as compared to those shown in Case 1, is the response of the controller when the circulating fluid mass flow rate  through reactor A saturates at its upper bound of 0.8 kg/s as shown at $t=80$ min in Figure ~\ref{fig:mdotA_cntr_heat_case3}. The MPC, recognizing this constraint, begins to increase the pressure differential across the compressor, since this is now the best input variable to use to control the heat transfer rate in Reactor A (see Figure~\ref{fig:dP_cntr_heat_case3}). Figure~\ref{fig:heatA_cntr_heat_case3} shows that the compressor is able to track the reference using the compressor pressure difference, but that the system is slower to respond given that the compressor pressure difference has a dynamic, rather than algebraic, effect on the output.  While Reactor A requires a larger mass flow rate over time because the reactor temperature is approaching the circulating fluid temperature, in Reactor B, the reactor temperature diverges from the circulating fluid temperature. Thus, as shown in Figure~\ref{fig:mdotB_cntr_heat_case3}, the mass flow rate required to meet the reference heat transfer rate decreases over time, so much so that that the input mass flow rate values used to achieve the final reference heat transfer rate are similar to the values used to achieve the initial value, albeit the final reference value is much larger.

\par Given the coupling of the dynamics between the two reactors, it is desirable to operate the system in a near-equilibrium state. This avoids the situation in which the compressor pressure difference and the mass flow rate in one reactor are saturated, which would degrade the performance of the controller. Near equilibrium, the absorption or desorption rate in each reactor balances the mass flow rate between them, and the energy transfer from the reaction balances the heat transfer to the circulating fluid in each reactor. It is therefore important to select a ratio between the reference heat transfer rates that allows for near-equilibrium operation if the controller is being used for a full charging or discharging cycle.

\subsection{Disturbance Rejection}
\par
\begin{figure}[!htb]
     \centering
     \begin{subfigure}{0.49\textwidth}
         \centering
         \includegraphics[width=\textwidth]{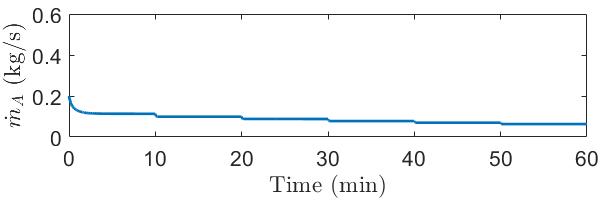}
         \caption{Circulating Fluid Mass Flow Rate A}
         \label{fig:mdotA_cntr_dist_case1}
         \vspace*{2.5mm}
     \end{subfigure}
     \begin{subfigure}{0.49\textwidth}
         \centering
         \includegraphics[width=\textwidth]{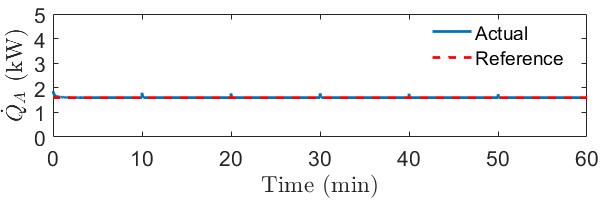}
         \caption{Heat Transfer Rate in Reactor A}
         \label{fig:heatA_cntr_dist_case1}
         \vspace*{2.5mm}
     \end{subfigure}
     \begin{subfigure}{0.49\textwidth}
         \centering
         \includegraphics[width=\textwidth]{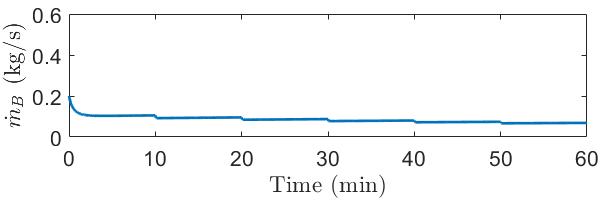}
         \caption{Circulating Fluid Mass Flow Rate B}
         \label{fig:mdotB_cntr_dist_case1}
         \vspace*{2.5mm}
     \end{subfigure}
     \begin{subfigure}{0.49\textwidth}
         \centering
         \includegraphics[width=\textwidth]{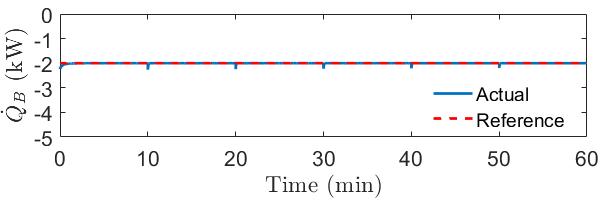}
         \caption{Heat Transfer Rate in Reactor B}
         \label{fig:heatB_cntr_dist_case1}
         \vspace*{2.5mm}
     \end{subfigure}
     \begin{subfigure}{0.49\textwidth}
         \centering
         \includegraphics[width=\textwidth]{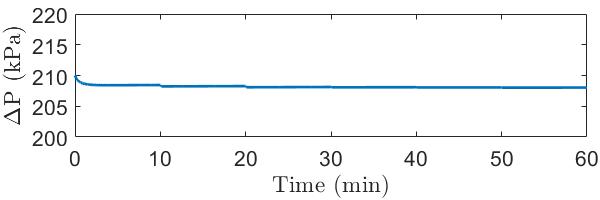}
         \caption{Compressor Pressure Difference}
         \label{fig:dP_cntr_dist_case1}
     \end{subfigure}
     \begin{subfigure}{0.49\textwidth}
         \centering
         \includegraphics[width=\textwidth]{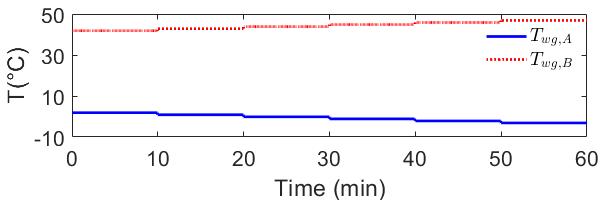}
         \caption{Circulating Fluid Temperatures}
         \label{fig:Twg_cntr_dist_case1}
     \end{subfigure}
     \caption{Input values set by the controller and output heat transfer rates in the hydride reactors for the case where hydrogen is being moved from reactor B to reactor A, compared to their reference values. The circulating fluid temperature changes every 10 minutes and the model is re-linearized every 30 minutes.}
    \label{fig:cntr_charge_changeT}
\end{figure}

 To test the controller's ability to mitigate exogenous disturbances, we consider constant heat transfer rate references in each reactor for the same sets of initial conditions described in the previous section, but now change the circulating fluid temperatures every 10 minutes as shown in Figure~\ref{fig:Twg_cntr_dist_case1}. For this case, the model is still re-linearized every 30 minutes. The control input signals and resulting heat transfer rates are compared to their reference values in Figure~\ref{fig:cntr_charge_changeT}.

\par
\begin{figure}[!htb]
     \centering
     \begin{subfigure}{0.49\textwidth}
         \centering
         \includegraphics[width=\textwidth]{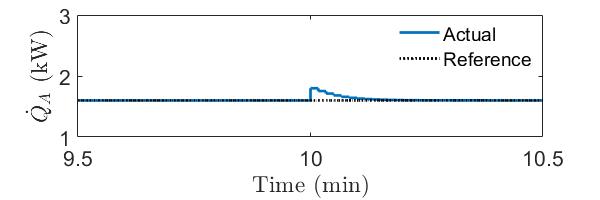}
         \caption{Heat Transfer Rate A}
         \label{fig:heatA_case1_zoomed}
     \end{subfigure}
     \begin{subfigure}{0.49\textwidth}
         \centering
         \includegraphics[width=\textwidth]{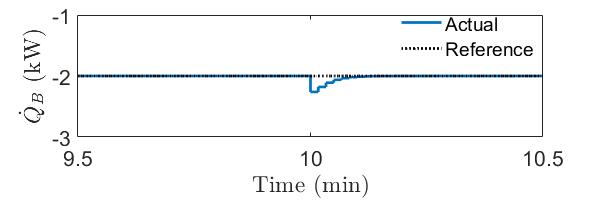}
         \caption{Heat Transfer Rate B}
         \label{fig:heatB_case1_zoomed}
     \end{subfigure}
     \caption{Heat transfer rates in the reactors for Case 1 from 9.5 to 10.5 minutes.}
     \label{fig:heat_case1_zoomed}
\end{figure}

 As shown in Figure~\ref{fig:cntr_charge_changeT}, the controller quickly adjusts to the changes in the disturbance inputs, successfully minimizing regulation error after only a very brief deviation away from the desired heat transfer rate. The first of these spikes can be seen in more detail in Figure~\ref{fig:heat_case1_zoomed}, which shows the $1$-minute period of time around the change in the disturbance inputs. From this, we can see that the controller converges to the reference value within 5-6 seconds. As with the case in which we considered time-varying heat transfer rates, the controller primarily changes the mass flow rates in order to track the reference, but does make some small changes to the compressor pressure difference. The controller is robust to changes in the disturbance inputs despite using a prediction model that is re-linearized less frequently than the disturbance signals change.

In a further test of the ability of the controller to mitigate changes in the disturbance inputs, the controller was simulated for the initial conditions given in Case 2, with a larger change in the water glycol temperatures. Results for this case are given in~\ref{sec:appB}.

%% file: 04_Conclusion.tex
\section{Conclusion}
\par In this paper we presented a a model predictive controller (MPC) for a two-reactor metal hydride system in which reactions in each hydride bed are driven by heat transfer between the metal hydride and a circulating fluid as well as a compressor moving hydrogen between the reactors. The multivariable controller successfully tracks desired values for the heat transfer between the hydride bed and the circulating fluid in each reactor by controlling the pressure difference produced by the compressor and the mass flow rates of the circulating fluid in each reactor. We derived a first-principles nonlinear dynamic model of the two-reactor system, and then linearized it for the purposes of control design. While the nonlinear model of the reactor neglects any temperature or pressure gradients within the hydride beds, it predicts the evolution of the reactor pressure, temperature, and heat transfer rates to the circulating fluid. By analytically linearizing the nonlinear model, we obtained a parameterized state-space model that could be easily updated for different operating conditions. This was leveraged in the design of the MPC, in which the linear prediction model was periodically re-linearized to minimize differences between its predictions and the dynamics of the nonlinear model. Through a series of simulated case studies, we demonstrated the performance of the controller for both reference tracking and disturbance rejection. The proposed model and multivariable controller enable continuous operation of two-reactor metal hydride systems to regulate heat transfer in a metal hydride energy storage or A/C system. Areas of future research include examining the viability of controlling heat transfer in one reactor while minimizing temperature change in both, as well as testing the controller for use with higher-order models and real hydride systems.


%% file: 05_Acknowledgements.tex
\section*{Acknowledgements}
\par Funding for this project was provided by the Center for High-Performance Buildings at Purdue University.

\section*{Model Availability}
\par The code used to linearize the nonlinear dynamics model is available as Matlab .m files on Github at https://github.com/patrickkrane/hydride-linearization-model.git.

%% file: 06_Appendix.tex
\appendix
\section{Controller Reference Tracking Results for Simulation Case 2}
\label{sec:appA}

\begin{figure}[!htb]
     \centering
     \begin{subfigure}{0.49\textwidth}
         \centering
         \includegraphics[width=\textwidth]{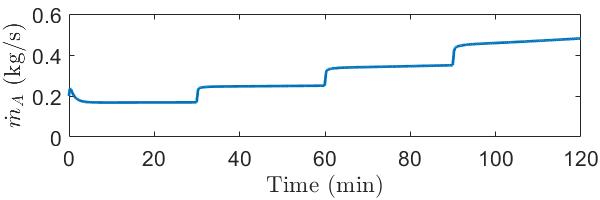}
         \caption{Circulating Fluid Mass Flow Rate A}
         \label{fig:mdotA_cntr_heat_case2}
         \vspace*{2.5mm}
     \end{subfigure}
     \begin{subfigure}{0.49\textwidth}
         \centering
         \includegraphics[width=\textwidth]{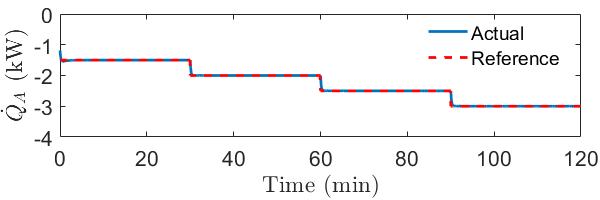}
         \caption{Heat Transfer Rate in Reactor A}
         \label{fig:heatA_cntr_heat_case2}
         \vspace*{2.5mm}
     \end{subfigure}
     \begin{subfigure}{0.49\textwidth}
         \centering
         \includegraphics[width=\textwidth]{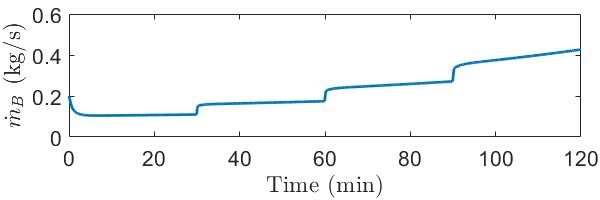}
         \caption{Circulating Fluid Mass Flow Rate B}
         \label{fig:mdotB_cntr_heat_case2}
         \vspace*{2.5mm}
     \end{subfigure}
     \begin{subfigure}{0.49\textwidth}
         \centering
         \includegraphics[width=\textwidth]{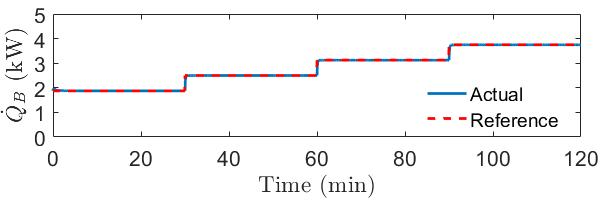}
         \caption{Heat Transfer Rate in Reactor B}
         \label{fig:heatB_cntr_heat_case2}
         \vspace*{2.5mm}
     \end{subfigure}
     \begin{subfigure}{0.49\textwidth}
         \centering
         \includegraphics[width=\textwidth]{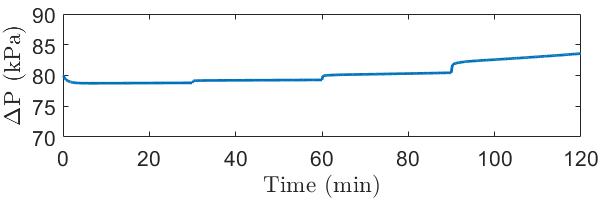}
         \caption{Compressor Pressure Difference}
         \label{fig:dP_cntr_heat_case2}
     \end{subfigure}
     \caption{Input values set by the controller and output heat transfer rate in each hydride reactor compared to the reference values for the case in which hydrogen is being moved from reactor A to reactor B. The model is re-linearized and the reference values changed every 30 minutes.}
    \label{fig:cntr_discharge_changeQ}
\end{figure}

\par Figures~\ref{fig:heatA_cntr_heat_case2} and~\ref{fig:heatB_cntr_heat_case2} demonstrate that the controller is again able to track the desired heat transfer rates in the reactors. The input values used to achieve this result are shown in Figures~\ref{fig:mdotA_cntr_heat_case2},~\ref{fig:mdotB_cntr_heat_case2},and~\ref{fig:dP_cntr_heat_case2}. As in Case 1, the controller successfully tracks the reference values. From Figures~\ref{fig:mdotA_cntr_heat_case2} and~\ref{fig:mdotB_cntr_heat_case2}, we see that the circulating fluid mass flow rates increase as needed to track the time-varying references. Finally, from Figure~\ref{fig:dP_cntr_heat_case2}, we see that the controller does not use the compressor pressure difference as much as the circulating fluid mass flow rates to achieve the desired tracking performance.

\section{Disturbance Rejection Results for Simulation Case 2}
\label{sec:appB}

\par
\begin{figure}[!htb]
     \centering
     \begin{subfigure}{0.49\textwidth}
         \centering
         \includegraphics[width=\textwidth]{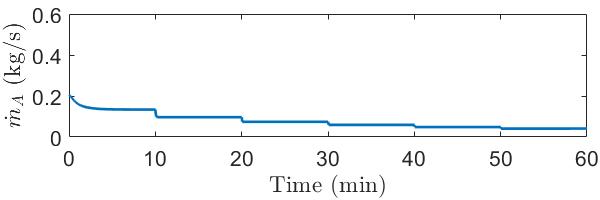}
         \caption{Circulating Fluid Mass Flow Rate A}
         \label{fig:mdotA_cntr_dist_case2}
         \vspace*{2.5mm}
     \end{subfigure}
     \begin{subfigure}{0.49\textwidth}
         \centering
         \includegraphics[width=\textwidth]{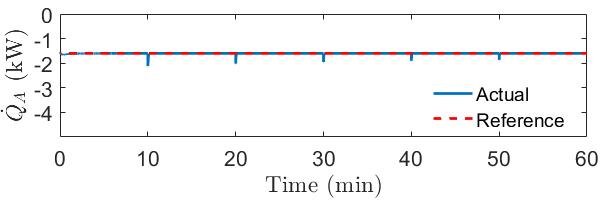}
         \caption{Heat Transfer Rate in Reactor A}
         \label{fig:heatA_cntr_dist_case2}
         \vspace*{2.5mm}
     \end{subfigure}
     \begin{subfigure}{0.49\textwidth}
         \centering
         \includegraphics[width=\textwidth]{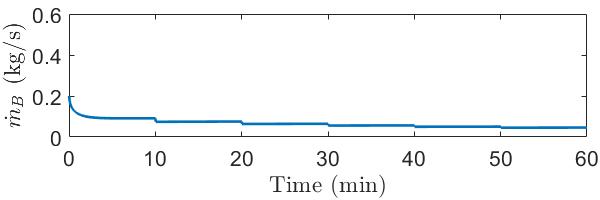}
         \caption{Circulating Fluid Mass Flow Rate B}
         \label{fig:mdotB_cntr_dist_case2}
         \vspace*{2.5mm}
     \end{subfigure}
     \begin{subfigure}{0.49\textwidth}
         \centering
         \includegraphics[width=\textwidth]{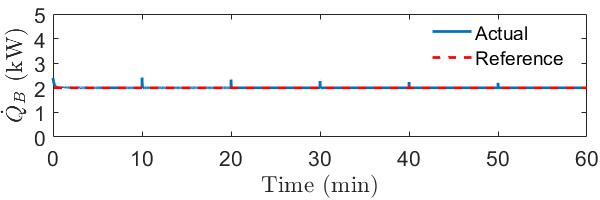}
         \caption{Heat Transfer Rate in Reactor B}
         \label{fig:heatB_cntr_dist_case2}
         \vspace*{2.5mm}
     \end{subfigure}
     \begin{subfigure}{0.49\textwidth}
         \centering
         \includegraphics[width=\textwidth]{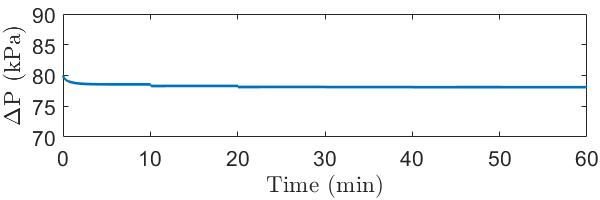}
         \caption{Compressor Pressure Difference}
         \label{fig:dP_cntr_dist_case2}
     \end{subfigure}
     \begin{subfigure}{0.49\textwidth}
         \centering
         \includegraphics[width=\textwidth]{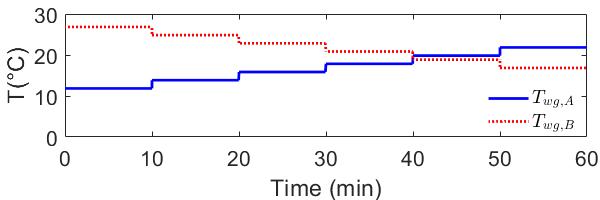}
         \caption{Circulating Fluid Temperatures}
         \label{fig:Twg_cntr_dist_case2}
     \end{subfigure}
     \caption{Input values set by the controller, output heat transfer rates in the hydride reactors, and changing disturbance inputs for the case in which hydrogen is being moved from reactor A to reactor B. The circulating fluid temperature changes every 10 minutes and the model is re-linearized every 30 minutes.}
    \label{fig:cntr_discharge_changeT}
\end{figure}

 To further evaluate the disturbance rejection capability of the controller, the change in the disturbance inputs is increased in Case 2. The water glycol temperatures for this case are given in Figure~\ref{fig:Twg_cntr_dist_case2}. The control inputs and the resulting outputs, compared to the reference values, are shown in Figure~\ref{fig:cntr_discharge_changeT}.

\par As shown in Figure~\ref{fig:cntr_discharge_changeT}, the controller still regulates the heat transfer rates to the desired values with only brief deviations in the regulation error (away from zero) whenever the disturbance inputs change. These results are again achieved primarily by varying the mass flow rates, with only slight use of the compressor. Since the changes in the circulating fluid temperatures are larger in this case, larger changes in the mass flow rates are needed to continue regulating the heat transfer rates to their references, as can be seen by comparing Figures~\ref{fig:mdotA_cntr_dist_case1} and~\ref{fig:mdotB_cntr_dist_case1} to Figures~\ref{fig:mdotA_cntr_dist_case2} and~\ref{fig:mdotB_cntr_dist_case2}. 